\documentclass[fleqn,usenatbib]{mnras}
\usepackage{newtxtext,newtxmath}
\usepackage{comment}
\usepackage{multirow}
\usepackage{mathtools}
\usepackage[T1]{fontenc}    
\usepackage{ae,aecompl}

\usepackage{graphicx}	
\usepackage{amsmath}	
\usepackage[ruled,vlined]{algorithm2e}
\usepackage{algpseudocode}
\usepackage{enumitem} 
\usepackage{xfrac} 
\usepackage[table]{xcolor}

\newcommand{\mdot}{$\mathrm{M}_\odot$}
\newcommand{\zdot}{$\mathrm{Z}_\odot$}

\newcommand{\sphgal}{\textsc{sphgal}}
\newcommand{\gadget}{\textsc{gadget-3}}
\newcommand{\healpix}{\textsc{healpix}}
\newcommand{\treecol}{\textsc{treecol}}

\newcommand{\ketju}{\textsc{ketju}}

\newcommand{\boost}{\textsc{BoOST}}

\definecolor{lightred}{RGB}{255,205,205}
\definecolor{lightblue}{RGB}{205,225,255}
\newcommand{\hi}{\cellcolor{lightblue}}
\newcommand{\lo}{\cellcolor{lightred}}

\usepackage{ulem}
\usepackage{soul}

\title[Galaxy evolution with variable supernova energies]{Dwarf galaxy simulations with stellar-core mass- and metallicity-dependent core-collapse supernova explosion energies}
\author[Lahén et al.]{Natalia Lahén$^{1,2}$\thanks{E-mail: nlahen@mpa-garching.mpg.de}, Kiril Maltsev$^{3}$, Thorsten Naab$^{1}$, Hans-Thomas Janka$^{1}$\\
$^{1}$Max-Planck-Institut f\"ur Astrophysik, Karl-Schwarzschild-Str. 1, 
D-85748, Garching, Germany\\
$^{2}$Zentrum f\"ur Astronomie der Universit\"at Heidelberg, Astronomisches Rechen-Institut, M\"onchhofstr. 12-14, D-69120 Heidelberg\\
$^{3}$ Max-Planck-Institut für Astronomie, Königstuhl 17, D-69117 Heidelberg\\
}
\date{Accepted XXX. Received YYY; in original form ZZZ}
\pubyear{2026}

\begin{document}
\label{firstpage}
\pagerange{\pageref{firstpage}--\pageref{lastpage}}
\maketitle

\begin{abstract}
Core-collapse supernovae (CCSNe) are a critical source of thermal and kinetic energy and chemically enriched material in galaxies. Leveraging recent advances in CCSN modelling, we implement a novel core-mass- and metallicity-dependent prescription for stellar explodability and CCSN explosion energies, $E_\mathrm{expl}$, in the SPHGal galaxy evolution code. It accounts for black hole (BH) formation via SN failures and via partial fallback of ejecta mass, and adopts an explosion-progenitor connection calibrated to three-dimensional CCSN simulations, such that $E_\mathrm{expl}$ varies between $<$\,0.1--2\,B. We ran simulations of isolated low-metallicity galaxies with variations in the input stellar feedback models. With the new explodability prescription nearly three times more BHs form compared to the canonical CCSN-model, but the reduced number of successful CCSNe has minimal impact on galactic star formation owing to the low fraction of failed CCSNe. While BH formation contributes to removal of matter from the galactic baryon cycle, the latter remains driven by galactic outflows, and most compact remnant mass is stored in white dwarves. The most significant change is instead brought about by the new average-$E_\mathrm{expl}$ that amounts to \mbox{$\sim$33\%} of the \mbox{$1\,\mathrm{B}=10^{51}$\,erg} traditionally assumed in sub-resolution CCSN-models. This reduction results in increased and less bursty star formation, reduced outflows and mass-loading factors, increased star cluster formation efficiency, and enhanced chemical enrichment and metal-recycling in star-forming regions. These results stand in contrast with outflow-launching and baryon cycle modelling in lower-resolution galaxy simulations in which canonical (i.e., 1\,B) CCSN feedback is typically assumed to produce galaxy models that match observed scaling relations.
\end{abstract}



\section{Introduction}

Stellar feedback is a fundamental ingredient in the regulation of the interstellar matter cycle and star formation especially in low-mass dwarf galaxies \citep{2011MNRAS.417..950H, 2020MNRAS.491.1656A}. Galactic outflows, driven by either evolving stars or accreting massive black holes (BHs), are considered the main astrophysical processes to producing realistic galaxies in cosmological structure formation models \citep[see e.g.][]{2015ARA&A..53...51S,2017ARA&A..55...59N}. Supernova (SN) explosions stir the interstellar medium (ISM) and maintain the hot phase subsequently launching outflows, accompanied with episodes of chemical enrichment  \citep[e.g.][]{2011ApJ...729...16K, 2016MNRAS.456.3432G,2016MNRAS.458.3528H, 2017MNRAS.466.1903G, 2018MNRAS.481.3325F, 2020MNRAS.495.1035S}. Before SNe occur, the early feedback processes, such as photoionizing radiation and stellar winds, pre-process the ISM for the disruptive effects of the SNe \citep[e.g.][]{2009ApJ...694L..26G, 2012MNRAS.424..377D,2012MNRAS.427..625W, 2013MNRAS.428..129S, 2013ApJ...770...25A,2014MNRAS.445..581H, 2017MNRAS.471.2151H,2018ApJ...859...68K}. While the early injection of pre-SN feedback in star-forming regions promotes the launching of the outflows, the combined effect may be reduction of final masses of star clusters \citep{2021MNRAS.506.3882S, 2021MNRAS.504.1039R,2022MNRAS.509.5938H, 2024A&A...681A..28A}. The consequence is a lower number of clustered SNe, which may alternatively decrease the collective impact of SNe on galactic scales.

In terms of the global energy budget and self-regulation within the galactic matter cycle, the contribution of core-collapse supernovae (CCSN) is particularly central. Because they originate from massive stars, they typically occur only a few million years after the onset of a star formation episode. Depending on the exact timing and local conditions, they can significantly impact their environment by heating and expelling any remaining gas from the star-forming region \citep{2017MNRAS.471.2151H,2021MNRAS.504.1039R,2026MNRAS.549ag959S}. 

However, there are several observations suggesting that when massive stars undergo iron-core collapse at the end of their lives, not all of them succeed to explode. Supporting clues are, for example, the lack of high-mass red supergiant progenitors to Type II SNe \citep{Kochanek2008, 2009MNRAS.395.1409S, Kochanek2014, 2015PASA...32...16S} (hereafter: \textit{RSG problem}); the mismatch between massive star formation and CCSN explosion rates \citep{Horiuchi2011}; electromagnetic transients followed by the disappearance of the stellar progenitor in the optical, consistent with a failing SN interpretation \citep{Gerke2015, De2026}; the successes in explaining the peak-trough distribution of chirp masses (low-mass end) of gravitational wave detected binary BH mergers 
via failed explosions of massive stars undergoing isolated binary evolution \citep{Schneider2023}; and the inferred negligible natal kick velocity of a BH in a binary with a luminous companion star \citep{Vigna-Gomez2024}. 

Furthermore, comparison of observed SN light curves to SN explosion simulation models have revealed an order of magnitude scatter in the explosion energies, $E_\mathrm{expl}$, of Type II SNe \citep{2009ApJ...703.2205K}.
The complexity of CCSN outcomes thus stands in contrast with simplified modelling assumptions that are widely adopted in galaxy-scale simulations. Save for a few exceptions, the typical assumption in galaxy simulations is that all massive stars with zero-age-main-sequence mass, $M_\mathrm{ZAMS}$, above $\sim8$--$10$ \mdot{} and up to some limiting upper mass explode with $E_\mathrm{expl}= 10^{51} \,  \mathrm{erg}= 1\, \mathrm{B}$.
While \citet{2009ApJ...703.2205K} indeed infer a typical $E_\mathrm{expl}\sim 0.9 \, \mathrm{B}$, observations are biased toward more energetic events, which are brighter. A ``typical'' or ``mean''- $E_\mathrm{expl}$ is therefore difficult to estimate, but a more recent analysis suggests that the statistically dominant Type II SN explosions of $M_\mathrm{ZAMS}\sim 9-15 \, M_\odot$ progenitors -- while covering a range from 0.2 to 1.4 B -- most frequently have $E_\mathrm{expl} < 0.9 \, \mathrm{B}$ \citep{Martinez2022}. Detailed three-dimensional (3D) hydrodynamics simulations of iron-core collapse SN explosions of low-mass progenitors favour rather low ``typical'' $E_\mathrm{expl}$, e.g. $\sim 0.1 \, \mathrm{B}$ for 9 $M_\odot$ progenitors \citep{Stockinger2020, Burrows2024, Janka2024}.  Apart from SN failures, galaxy-scale SN feedback models could thus also benefit from accounting for progenitor-dependence of $E_\mathrm{expl}$ of CCSNe.

Disentangling the role and emergent effects of each stellar feedback channel on galaxy evolution requires detailed high-resolution simulations incorporating treatment of each mechanism, with the ability to toggle each on and off independently. The various feedback channels acting in concert have been addressed in hydrodynamical simulations in both cosmological \citep{2013ApJ...770...25A, 2020MNRAS.491.1656A, 2023MNRAS.519.3154H, 2026MNRAS.548ag375S, 2025OJAp....8E.153K, 2025arXiv251005201K} and idealised \citep{2016MNRAS.458.3528H, 2019MNRAS.482.1304E, 2021MNRAS.506.3882S, 2021PASJ...73.1036H, 2023MNRAS.521.2196A, 2024A&A...691A.231D} settings, both using stellar population averaged prescriptions and on individual star basis. Catalogs of massive star evolution up to the onset of iron-core infall and their model-dependent
 $E_\mathrm{expl}$-outcomes are available, and thus can be used for construction of SN feedback prescriptions in galaxy evolution codes. Examples are the models at solar metallicity ($Z=Z_\odot$), for instance, by \citet{2016ApJ...821...38S}, \citet{2016ApJ...818..124E}, \citet{2019ApJ...870....1E} and \citet{2020ApJ...890...51E} and at lower $Z$ by \citet{2020ApJ...888...91E}. 
Previous studies by \cite{2021MNRAS.501.5597G} and \citet{2025ApJ...979...44S} in particular used tabulated results of CCSN explosion outcomes from \citet{2016ApJ...821...38S} computed at near-solar $Z$ to perform low-mass galaxy simulations. When convolved with the initial stellar mass function (IMF), they found that the combined effect of progenitor-variable explodability and $E_\mathrm{expl}$ is to reduce the cumulative CCSN energy available to generate superbubbles and to drive outflows. In those simulations, the reduced CCSN energy resulted in increased star formation rates and reduced galactic outflows.

In this paper we focus on the impact of stellar progenitor dependence of SN explodability and $E_\mathrm{expl}$ on star-forming galaxies at low $Z$. As an alternative to looking up the SN outcome from a tabulated set of explosion properties at fixed $Z$ and $M_\mathrm{ZAMS}$, here we establish a link between massive star evolution models and CCSN outcomes using the final carbon-oxygen (CO) core mass, $M_\mathrm{CO}$, and a continuous $Z$-dependence. We study the effects of a statistical CCSN model predicting explodability outcomes (failed vs successful SN) and compact remnants (neutron star [NS] vs BH) as a function of a star's $M_\mathrm{CO}$ and $Z$,  introduced in \citet{2025A&A...700A..20M} (hereafter: \textit{M25}). The model has recently been extended to predict $E_\mathrm{expl}$ of the exploding CCSNe (Maltsev et al., in prep.; hereafter: \textit{M25, variable E}).  

We implement the \textit{M25, variable E} model in the hydrodynamical galaxy evolution code \textsc{sphgal} \citep{2014MNRAS.443.1173H, 2016MNRAS.458.3528H, 2017MNRAS.471.2151H, 2023MNRAS.522.3092L} to study the evolution of a low-$Z$ dwarf galaxy. The simulations are part of the Galaxy Realizations Including Feedback From INdividual massive stars (\textsc{griffin})\footnote{\url{https://wwwmpa.mpa-garching.mpg.de/~naab/griffin-project}} project, extending the suite of solar-mass resolution galaxy evolution models that explore e.g. clustered star formation \citep{2020ApJ...891....2L, 2022MNRAS.509.5938H, 2023MNRAS.522.3092L}, the cold cloud life-cycle \citep{2024MNRAS.534..215F} as well as BH evolution \citep{2025MNRAS.537..956P}. We utilize the CO core masses available in the Bonn Optimized Stellar Tracks (\boost; \citealt{2022A&A...658A.125S}) single-star model data base, which has previously been implemented in \sphgal{}. Mapping the catalog of stellar evolution to the \textit{M25, variable E} model via $M_\mathrm{CO}$ and $Z$, we obtain predictions of CCSN explosions with $E_\mathrm{expl}$ varying from $<0.1$ to $2 \, $ B and BH formation either via partial fallback of SN ejecta mass or via failed SNe.
We test the effect of the new CCSN model against the previous implementation that assumed a fixed $E_\mathrm{expl}= \mbox{1 B}$ and compare the results to additional variations of other feedback channels e.g. by running a simulation with overall reduced CCSN energy or by switching the photoionizing radiation completely off.

This paper is organized as follows. Section \ref{section:methods} provides a description of the simulation method, including the new CCSN explodability and $E_\mathrm{expl}$-prescription explored here given in Section \ref{section:CCSN_model}. Section \ref{section:results} gives results of the impact of CCSN model variation on isolated dwarf galaxy evolution on a timescale of 1 Gyr, comparing  star formation rates, CCSN environments and star cluster properties. Section \ref{section:discussion} discusses these upon the background of observational constraints, and section \ref{section:conclusions} provides conclusions.

\section{Simulations}\label{section:methods}

For hydrodynamical simulations of dwarf galaxy evolution, we use the smoothed particle hydrodynamics (SPH) code \sphgal{} \citep{2014MNRAS.443.1173H, 2016MNRAS.458.3528H, 2017MNRAS.471.2151H}, which is a modified version of \gadget{} \citep{2005MNRAS.364.1105S}. In addition we use modifications for star formation and stellar feedback introduced in \citet{2019MNRAS.483.3363H} and \citet{2023MNRAS.522.3092L} and further develop the methods for CCSNe as outlined below.

\subsection{Thermochemistry and star formation}

As described in \citet{2016MNRAS.458.3528H, 2017MNRAS.471.2151H}, the heating and cooling processes in the low-temperature regime
($<3\times 10^4$ K) are computed using a chemical network (H$_2$, H$^+$, H, CO, C$^+$, O and free electrons) following the methods of \citet{1997ApJ...482..796N}, \citet{2007ApJS..169..239G} and \citet{2012MNRAS.421..116G}. For the cooling rate of higher temperature gas ($>3\times 10^4$ K) we use the tabulated metallicity dependent rates from \citet{2009MNRAS.393...99W}. Dust is present at a constant gas mass fraction of $0.1\times Z\sim 10^{-3}$.  

We adopt the star formation prescription from \citet{2019MNRAS.483.3363H} and \citet{2023MNRAS.522.3092L} but exclude the fully sampled stellar mass function to run the simulations for \mbox{1 Gyr} at a reduced computational cost. Gas particles with mass $m_\mathrm{SPH}\approx 4$ \mdot{} are converted into star particles when the local SPH kernel mass $M_\mathrm{SPH}=100 \, m_\mathrm{SPH}$ exceeds the Jeans-mass \begin{equation}
   M_\mathrm{J} = \frac{\pi^{5/2}c_\mathrm{s}^3}{6G^{3/2}\rho^{1/2}}
\end{equation}
by $M_\mathrm{J}<0.5 \, M_\mathrm{SPH}$, where $c_\mathrm{s}$ is the sound speed, $\rho$ the gas density and $G$ is the gravitational constant. We approximate the timescale of gravitational collapse of the gas cloud with the free-fall time, $t_\mathrm{dyn}=(4\pi G \rho)^{-1/2}$, given by the gas density at star formation before sampling the converted mass into a distribution of initial stellar masses. After the delay the stellar mass available within the local Jeans length
\begin{equation}
   R_\mathrm{J} = \left( \frac{3}{4\pi} \frac{M_\mathrm{J}}{\rho} \right) ^{1/3},
\end{equation}
is sampled according to the \citet{2001MNRAS.322..231K} mass function between \mbox{0.08 \mdot{}} and \mbox{500 \mdot} and stored in the particle data from \mbox{1 \mdot{}} upward. Lower mass stars only contribute to the total stellar mass stored in each particle. If there is not enough mass within the Jeans length for spawning a randomly drawn stellar mass, the value is discarded and a new value drawn for another attempt. The simulations here consider quiescent low-metallicity dwarf galaxies and the local reservoir available for IMF sampling remains typically low enough that successful creation of stars exceeding \mbox{$M_\mathrm{ZAMS}>100$ \mdot{}} is extremely rare.

\subsection{Stellar feedback via radiation and winds}

The methods from \citet{2016MNRAS.458.3528H, 2017MNRAS.471.2151H} and \citet{2023MNRAS.522.3092L} are used for energy and mass injection through stellar winds, photoionizing radiation and far-ultraviolet radiation. The interstellar radiation field is determined by the flux at \mbox{$6$--$13.6$ eV}, attenuated using the column density computed within the shielding length of \mbox{50 pc} using the \treecol-algorithm \citep{2012MNRAS.420..745C} in 12 solid angles divided using the \healpix-algorithm \citep{2011ascl.soft07018G} around the location of each gas particle. Massive stars ($>9$ \mdot{}) are given hydrogen ionizing photon rates at $>13.6$ eV and far-ultraviolet fluxes calculated based on the Bonn Optimized Stellar Tracks (\boost; \citealt{2022A&A...658A.125S}) at corresponding initial metallicity. For far-ultraviolet flux of low-mass stars (\mbox{$<9$ \mdot}) we use the \textsc{BaSeL} spectral library at $Z\sim 0.1$ \zdot{} \citep{1997A&AS..125..229L, 1998A&AS..130...65L, 2002A&A...381..524W} and the \textsc{geneva} models at the corresponding metallicities \citep{2013A&A...558A.103G}. For the mass $M_\mathrm{wind}$ expelled in stellar winds (continuous, stellar age dependent momentum and mass injection) of massive stars we likewise use the \boost{} tables. The final mass of each massive star before the evaluation of SN outcome is then $M_\mathrm{final}=M_\mathrm{ZAMS}-M_\mathrm{wind}$, which includes the He and CO core masses and an envelope prescribed as the chemical composition at the end of He-burning. Stars with initial masses between $0.5$--\mbox{$8$ \mdot{}} release a burst of asymptotic giant branch winds \citep{2010MNRAS.403.1413K} at the end of their lives. Stellar lifetimes are approximated either using the Geneva models or the \boost{} models depending on the initial mass. Galactic chemical enrichment proceeds through stellar evolution and we track the mass of 13 chemical elements per particle: H, He, N, C, O, Si, Al, Na, Mg, Fe, S, Ca and Ne.

\begin{figure*}
\includegraphics[width=\textwidth]{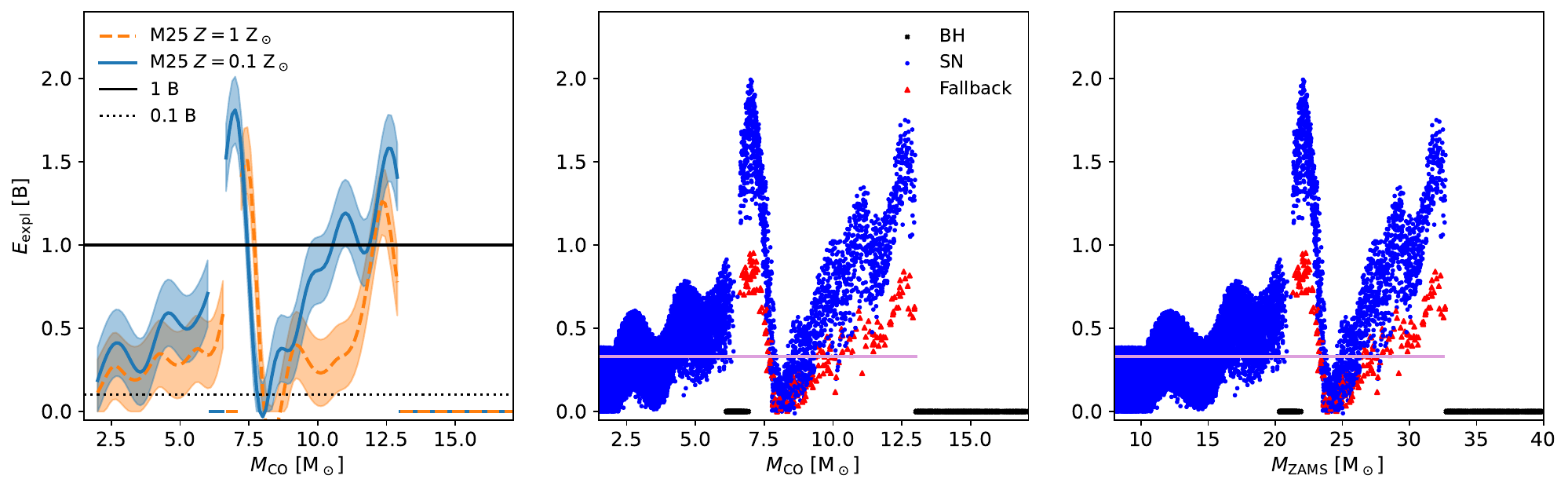}
\caption{Comparison of the $E_\mathrm{expl}$-landscape, predicted by the  \textit{M25, variable E} and the \textit{Original} CCSN models employed in the dwarf galaxy simulation, as a function of CO core mass (left and middle panel) and initial stellar mass (right panel) between $8\,\mathrm{M}_\odot<M_\mathrm{ZAMS}<40\,\mathrm{M}_\odot$. The curves in the \textit{left panel} show the \textit{M25, variable E}-predictions at 1 \zdot{} (dashed orange) and 0.1 \zdot{} (solid blue) that we use to interpolate and sample CCSN explosion energies at $Z$ in-between. The gaps in $E_\mathrm{expl}$-curves indicate failed explosions and BH formation. The horizontal lines indicate the progenitor-independent $E_\mathrm{expl}$-predictions -- of \mbox{1 B} (solid) and of \mbox{0.1 B} (dotted), respectively, in the \textit{Original} model. The data points in the \textit{middle} and \textit{right} panels show the successful (blue circles), failed (black crosses) and fallback (50\% of the energy and mass of a respective successful CCSN for 10\% of CCSN events between the failed regimes; red triangles) CCSN events in the dwarf galaxy simulation run using the full metal-dependent explodability model with scatter, starting with initial $Z=0.1$ \zdot. The horizontal lines here indicate the mean realized energy of successful CCSNe in this simulation. The gap of zero explodability around $M_\mathrm{CO}\sim 6.5$ is not completely free of events in the simulation because the metallicity of newly forming stars is evolving. \label{fig:explosions}}
\end{figure*}

\subsection{Pair-Instability supernovae}\label{section:PISN}
We divide the SN outcomes of massive stars into two categories: 1) neutrino-driven CCSNe and 2) pair-instability supernovae (PISNe). Our models for CCSNe are described in detail in Sect. \ref{section:CCSN_model}, and their effects on galaxy evolution are in the focus of this work. For PISNe, we assume that these occur for stars that reach a helium core mass between $65$--\mbox{$133$ \mdot{}}, with yields and explosion energies ($5$ -- $90$ B) from \citet{2002ApJ...567..532H}. PISN events are thought to be disintegrating the entire star leaving no compact remnant behind\footnote{We neglect the BH forming pulsational PISNe here} but are extremely rare in the simulated quiescent galaxies where stars seldom form in dense enough regions to sample the high-mass tail of the IMF that reaches up to the PISN range (see \citealt{2023MNRAS.522.3092L}). Overall the simulations produce less than ten PISNe when using the fiducial feedback models described here, and of the order of 50 PISNe when photoionizing feedback is turned off. PISNe are completely switched off in the one simulation where we test an intentionally low CCSN energy of 0.1 B.

\begin{table*}
\begin{tabular}{l c c c l}
\hline
Name & CCSN progenitor $M_\mathrm{ZAMS}$ [\mdot] -range & $E_\mathrm{expl}$-range & Fallback-BH formation & Other changes\\
\hline
\texttt{M25, variable E} & $(8, 32.7)$ & \mbox{0 B} (failed) to \mbox{$\sim 2$ B} & Yes & - \\
\texttt{M25, fixed E}    & $(8, 32.7)$ & \mbox{0 B} (failed) or \mbox{1 B} & Yes & - \\
\texttt{Original}        & $(8,40)$ & \mbox{1 B} & No & - \\
\texttt{Original, 0.1 B} & $(8,40)$ & \mbox{0.1 B} & No & No PISNe \\
\texttt{Original, no PI} & $(8,40)$ & \mbox{1 B} & No & No photoionization \\
\hline
\end{tabular}
\caption{Summary of the five CCSN models used in the dwarf galaxy simulation. The columns provide the name of the simulation, ZAMS mass range of the CCSNe, $E_\mathrm{expl}$ range, whether fallback BH formation is enabled, and other remarks.}
\label{table:parameters}
\end{table*}

\subsection{Iron-core collapse supernovae}\label{section:CCSN_model}

For CCSNe, we adopt two model classes: 1) the \texttt{Original} model, according to which all stars with $M_\mathrm{ZAMS}$ in-between $8$--$40$ \mdot{} explode with a fixed $E_\mathrm{expl}$ (the canonical 1 B, and $E_\mathrm{expl}=\mbox{0.1 B}$ as variation \texttt{Original, 0.1 B} thereof), and 2) \textit{M25} based models, with fixed $E_\mathrm{expl} = 1 \, \mathrm{B}$ (\texttt{M25, Fixed E}) and with progenitor-dependent $E_\mathrm{expl}$ (\texttt{M25, variable E}; see Table \ref{table:parameters}). The \textit{M25}-model for explodability of massive stars is introduced in \citet{2025A&A...700A..20M}, and its open-access Python implementation is available via \citet{kmaltsev_2025_15350140}. Its \textit{M25, variable E}-extension will be formally introduced in Maltsev et al. (in prep.) and be made publicly available.

In either case, we adopt the $M_\mathrm{ZAMS}$- and $Z$-dependent explosive nucleosynthesis yields of \citet{2004ApJ...608..405C}. To conserve mass when we also include the \boost{} stellar tables, we reduce the pure CCSN ejecta mass $\tilde{M}_\mathrm{ej}$ by the amount already expelled in stellar winds by scaling the yield in each element down by a total mass fraction $f_\mathrm{ej}=(\tilde{M}_\mathrm{ej}-M_\mathrm{wind})/\tilde{M}_\mathrm{ej}$. The yields are reduced in proportion to retain the enrichment pattern of the yield table. The resulting NS remnant mass $M_\mathrm{NS}=M_\mathrm{final}-M_\mathrm{ej}=M_\mathrm{final}-f_\mathrm{ej}*\tilde{M}_\mathrm{ej}$ then approximately matches with the values in \citet{2004ApJ...608..405C} when summed over all tracked elements.

\paragraph*{SN failures:} 
To decide whether a massive star undergoing iron-core collapse at the end of its life succeeds or fails to explode, \textit{M25} evaluates its final $M_\mathrm{CO}$ and $Z$, while distinguishing between hydrogen-rich envelope retaining SN progenitors and those that lost their hydrogen-rich envelope via binary mass transfer or episodic mass loss (such as during the luminous blue variable phase). In our galaxy simulations, since we use the \boost{} massive single-star models at variable $Z$ and no binary evolutionary models, we only use the \textit{M25} prescription for hydrogen-rich envelope retaining SN progenitors. \textit{M25} is calibrated to the SN outcomes predicted by the semi-analytical CCSN model from \citet{Müller2016} (hereafter: M16) with parameter choice as in \citet{Schneider2021}, when used to explode the $Z=Z_\odot$ \citep{Schneider2021} and the $Z=Z_\odot/10$ \citep{Schneider2023} 
SN progenitor models, respectively. Notably, it predicts explodability outcomes that follow characteristic trends with $M_\mathrm{CO}$ that are preserved with variable $Z$: massive stars are predicted to successfully explode, unless their $M_\mathrm{CO}$ is 1) within a narrow ($\sim 0.6 \, M_\odot$ width) window of failed SNe at intermediate core masses (centered at $\sim 6.3 \, M_\odot$ for $Z=Z_\odot/10$  and at $\sim 6.7 \, M_\odot$ for  $Z=Z_\odot$), or 2) above a high-core mass threshold (at $\sim13 \, M_\odot$) toward a domain of failed SN outcomes extending over the remaining progenitor core mass range. When coupled to our PISN model, this means an exception are stars with helium core mass within the PISN core mass gap as described in Section \ref{section:PISN}. For the generalized $Z$-dependence of the $M_\mathrm{CO}$ thresholds demarcating the intermediate-mass failed SN window and high-mass failed SN domain, the authors adopt a parametric $\log Z$-scaling law.

We assume that the resulting BH mass, $M_\mathrm{BH}$, is given by the final total mass, $M_\mathrm{final}$, of the non-exploding progenitor star:
\begin{equation}
    M_\mathrm{BH} = M_\mathrm{final}
\end{equation}

\paragraph*{Fallback-BH formation and SN ejecta masses:}
While SN failure is the dominant BH formation channel, \textit{M25} also statistically models BH formation via partial fallback of SN ejecta mass  onto the proto-NS in successful CCSN explosions, thereby pushing it beyond the stability limit.
The authors suggest two different fallback-BH formation prescription variants (\textit{Model A} and \textit{Model B}, with the \textit{A} variant as the suggested default; see \citet{2025A&A...700A..20M} for details on \textit{Model A}). In this study, for simplicity, we adopt the \textit{Model B} variant, according to which a massive star of given $Z$ and $M_\mathrm{CO}$ above the intermediate-mass failed SN window but below the high-mass failed SN domain has a 10\% probability of undergoing a SN explosion with partial ejecta mass fallback leading to BH formation. We dice out individual outcomes accordingly using a stochastic remnant type prescription.
Under these modelling assumptions, fallback-BHs in the \texttt{M25, variable E} simulation constitute $\sim20$\% of BH formation events from our IMF-weighted sample of stars between $M_\mathrm{ZAMS}=8$--$40$ \mdot, and 11.6\% in total out of all BHs with $M_\mathrm{ZAMS}>8$ \mdot.

We model the ejecta mass $M_\mathrm{ej}$ in such events as
\begin{equation}
\label{eq:mej}
    M_\mathrm{ej} = (M_\mathrm{final} - M_\mathrm{PNS})/2
\end{equation}
where $M_\mathrm{PNS}$ is the proto-NS mass, i.e. assuming that half of the stellar mantle mass above it is ejected while the other half falls back, leaving behind a BH remnant of mass $M_\mathrm{BH}=M_\mathrm{final}-M_\mathrm{ej}$. In cases of robust CCSN explosions, we look up NS remnant masses, $M_\mathrm{NS}$, using the tables from \citet{2004ApJ...608..405C}, and also use these to assign $M_\mathrm{PNS}$
in eq. (\ref{eq:mej}).

In fallback-BH formation cases, we reduce the \textit{M25, Variable E}-  predicted $E_\mathrm{expl}$ (see below for the prescription) by 50\% to model a marginal success of the CCSN explosion mechanism, though we note that this modelling assumption comes ad hoc and might need to be revised in future simulations. While CCSNe with fallback-BH outcomes constitute 11.6\% of all the BHs formed in the simulations, these amount to only $\sim 1 \%$ of all successful CCSN explosions in our IMF-weighted sample.  Therefore, the specifics of our fallback-BH $E_\mathrm{expl}$-modelling assumptions will not change the general results of SN feedback effects studied in this work.

Due to the wide window of successful SN explosions of high-$M_\mathrm{CO}$ CCSN progenitors and the rarity of fallback-BH formation outcomes in this regime, \textit{M25} results in a prominent dearth in the distribution of BH masses between $\sim10$--$\sim15 \, M_\odot$
(specifics of the outcomes depend on the adopted stellar evolution models;see e.g. \citet{Wagg2026} and \citet{He2026} for its recent use in studies of BH demographics in simulated Milky-Way like galaxies). This predicted BH formation landscape recently received support by gravitational wave observations of binary BH (BBH) mergers with the release of the latest (\texttt{GWTC-5}) data catalog available to date \citep{GWTC-5}. \citet{Willcox2025} show that the use of \textit{M25} in
simulations of BBH merger formation via isolated binary evolution results in a dip in the resulting BBH chirp mass distribution at $\sim 10 \, \mathrm{M_\odot}$, which withstands variations in the adopted binary evolution and cosmic star formation history models. The $\sim 10 \, \mathrm{M_\odot}$-dip is indeed observed, and several other CCSN models do not reproduce it under the same model variation conditions in the binary population synthesis setup. Furthermore, \textit{M25} partially addresses the \textit{RSG problem} by the SN failure of progenitors within the intermediate-mass range, is consistent with the failed SN interpretation of the \texttt{N6946-BH1} observation \citep{Gerke2015, Adams2017} -- while explaining the explosion of Type Ib / Ic SNe going off from progenitors with more massive CO cores as inferred from
nebular line spectroscopy of SN remnants and from SN ejecta masses \citep{2025A&A...700A..20M}.

\paragraph*{CCSN explosion energies:}
The \textit{M25, variable E} model emerging from Maltsev et al. (in prep.) is constructed as follows:

\begin{enumerate}
    \item Fitting the explosion energy, $\tilde{E}_\mathrm{expl}$, predicted by M16 as a function of $M_\mathrm{CO}$ of the single-star SN progenitor models from \citet{Schneider2021} and \citet{Schneider2023} at $Z_\odot$ and $Z_\odot/10$, respectively. 
    \item Introducing the failed SN regimes as in \textit{M25}, over which $\tilde{E}_\mathrm{expl} = 0 \, \mathrm{B}$.
    \item For the general $Z$-dependence, interpolating the $\tilde{E}_\mathrm{expl}-M_\mathrm{CO}$ relation at these two $Z$ grid points according to a $\log Z$-scaling law.
    \item Re-calibrating the explosion energy predictions, 
    \begin{equation}
        \tilde{E}_\mathrm{expl} \longrightarrow E_\mathrm{expl}
    \end{equation}
    for improved agreement with the explosion-progenitor connection as obtained from three-dimensional (3D) CCSN simulations. 
\end{enumerate}

The outcome is an algorithm predicting $E_\mathrm{expl}(M_\mathrm{CO},Z)$ of Type II / Ib SNe over a continuous parameter space in $M_\mathrm{CO}$ and $Z \in (Z_\odot/10, Z_\odot)$ (see left panel of Fig. \ref{fig:explosions}). 

For step 3.), Maltsev et al. (in prep.) compare the $\tilde{E}_\mathrm{expl}$-predictions, casted by M16, over the set of SN progenitors compiled in \citet{Janka2024} to their explosion energy outcomes as obtained from 3D CCSN simulations. They find that M16 systematically overpredicts the explosion energy compared to the 3D outcome. To account for it, they fit the overprediction ``error'' using a function, $f(.)$, subtract it from the initial $\tilde{E}_\mathrm{expl}$-prediction, and model the remaining  disagreement between 3D and M16 by a symmetric random component, $\varepsilon$, such that
\begin{equation}
    E_\mathrm{expl}(M_\mathrm{CO},Z) = f\left(\tilde{E}_\mathrm{expl}(M_\mathrm{CO}, Z)\right) + \varepsilon
\end{equation}
with $|\varepsilon| \leq  0.2 \, \mathrm{B}$. Finally, this makes $E_\mathrm{expl}(M_\mathrm{CO},Z)$ a prescription that follows deterministic trends with $M_\mathrm{CO}$ and $Z$, but which receive a random modulation. 

In practice, we store the two $M_\mathrm{CO}$--$f\left(\tilde{E}_\mathrm{expl}\right)$ relations, at $Z=1$ \zdot{} and $Z=0.1$ \zdot, respectively, over a grid at 100 equally spaced values of $M_\mathrm{CO}/M_\odot \in (2, 13)$. For a given input-$M_\mathrm{CO}$ and input-$Z$, we interpolate these two relations logarithmically in $Z$ and randomize the output by the $\varepsilon$-term. For $M_\mathrm{CO}$ below 2 \mdot{} we interpolate using the lowest $M_\mathrm{CO}=2$ \mdot{} grid points.

Fig. \ref{fig:explosions} illustrates the relation between $M_\mathrm{CO}$ and $f\left( \tilde{E}_\mathrm{expl} \right)$ arising from the input \textit{M25, variable E} model at $Z=1$ \zdot{} and $Z=0.1$ \zdot, respectively. The main deterministic trends are the following:
as $Z$ decreases, $E_\mathrm{expl}$ tends to be higher, with the $E_\mathrm{expl}$-peak shifting toward lower $M_\mathrm{CO}$. Massive stars with $M_\mathrm{CO}$ below the intermediate-mass failed SN window, which due to the IMF are the most frequent, explode with $E_\mathrm{expl}$ well below $1 \, \mathrm{B}$. The most energetic explosions, far exceeding $1 \, \mathrm{B}$, are obtained for massive stars with $M_\mathrm{CO}$-masses right above the failed SN window or right below the high-mass threshold to the domain of failed SN outcomes.

While we do not model the propagation of cosmic rays, we point out that a non-negligible fraction of $E_\mathrm{expl}$ is used up into acceleration of cosmic rays when the SN blast wave hits the circumstellar medium of the progenitor and a SN remnant forms. The amount of energy injected into the ISM is therefore given by $\alpha \cdot E_\mathrm{expl}$ rather than $E_\mathrm{expl}$, where $\alpha < 1$ is a reduction factor due to cosmic ray acceleration and other effects. In galaxy-scale simulations that model cosmic ray transport, $\alpha \simeq 0.9$ is often adopted (e.g. \citealt{2017MNRAS.465.4500P, 2018MNRAS.479.3042G, 2019MNRAS.488.3716C}).
In this work, we do not model cosmic ray acceleration and therefore neglect this reduction effect.

\paragraph*{Connection to massive star evolution models:}
\textit{M25} is best conceived as a model for a massive star's evolution through the advanced burning phases up to the CCSN outcome. The physical origin of final stellar structures at onset of iron-core infall that are difficult to explode is linked to the transition from a radiation- to a neutrino-dominated core-carbon (intermediate-mass failed SN window) or core-neon (high-mass domain of SN failures) burning phase \citep{Laplace2025, Schneider2023, 2025A&A...700A..20M}.  To map between \textit{M25}-predicted outcomes and the catalog of stellar evolution models adopted in the galaxy simulation, we use $M_\mathrm{CO}$ rather than $M_\mathrm{ZAMS}$ or the helium core mass. The advantage of $M_\mathrm{CO}$ is that 1) within the $Z$-range considered here, it remains almost constant from its formation after the end of core-helium burning throughout until iron-core collapse and that 2) being a proxy for the core temperature, it is -- along with the core-carbon mass fraction -- the main parameter 
pre-determining the late-phase evolution through the advanced burning phases \citep{Chieffi2020, Patton2020, Schneider2023, Laplace2025, 2025A&A...700A..20M}.

Coupling \boost{} stellar models,  which are evolved up to core-carbon exhaustion, to \textit{M25} and the nucleosynthesis outcomes from \citet{2004ApJ...608..405C}, we complete the SN feedback model. This coupling gives us $M_\mathrm{final}$ (determining $M_\mathrm{ej}$) and the stellar lifetimes, $\tau_\mathrm{SN}$ (determining the delay time for each CCSN explosion event), along with $E_\mathrm{expl}$. 

\begin{figure}
\includegraphics[width=\columnwidth]{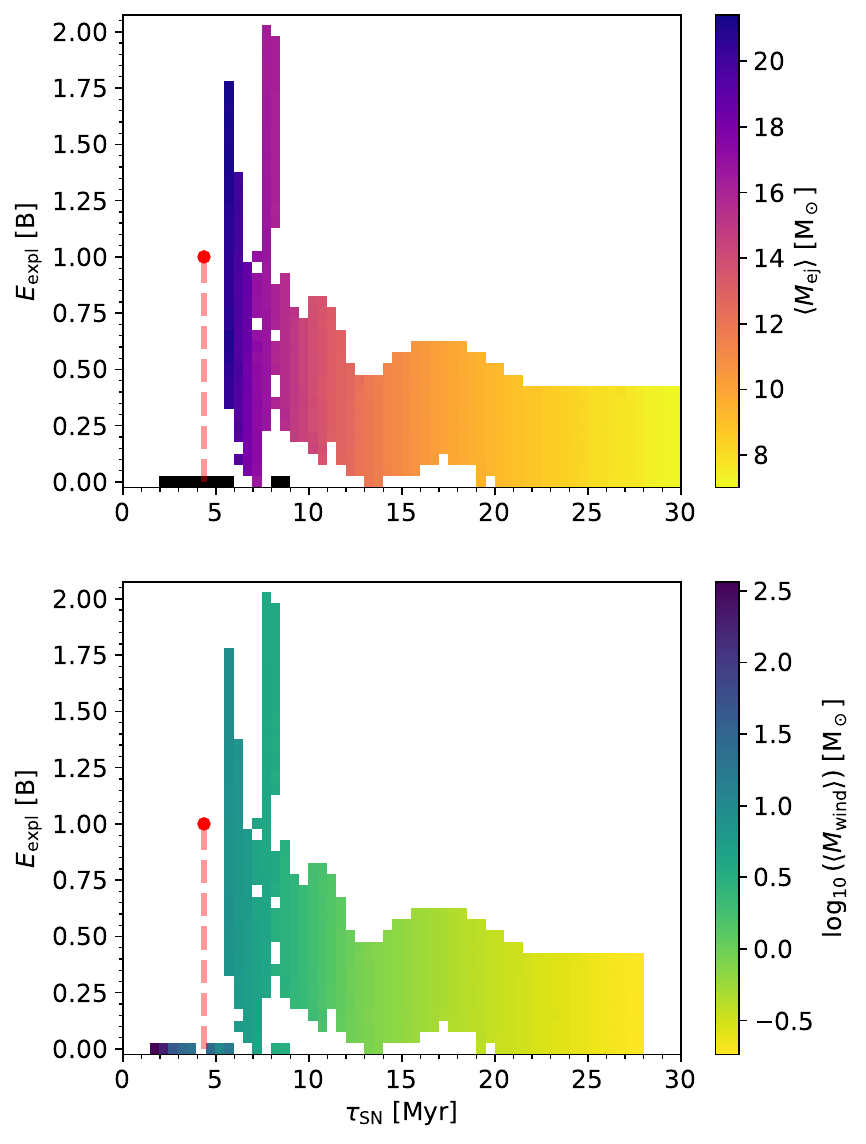}
\caption{The distribution of stellar lifetimes, $\tau_\mathrm{SN}$, and CCSN explosion energies, $E_\mathrm{expl}$, colour coded with the amount of mass ejected during stellar lifetimes via stellar winds (\textit{bottom})
and via the final explosion event (\textit{top}) in the \texttt{M25, variable E} CCSN model, when coupled to the \boost{} stellar progenitor model. Note the use of a logarithmic scale in the colour bar of the bottom panel, to visualize the large range of wind mass loss across three orders of magnitude in mass. The black pixels in the top panel indicate failed explosion outcomes for which the stellar wind ejecta can in rare cases range up to several $100$ \mdot. The red dot indicates the timing of the first CCSN in the \texttt{Original} run, occurring $\sim 1$--$2$ Myr earlier since it originates from a progenitor of $M_\mathrm{ZAMS}=$ \mbox{40 \mdot{}}.\label{fig:lifetimes}}
\end{figure}

Under these modelling assumptions, CCSN explosions occur with delay times given by the stellar lifetimes in the range of \mbox{$\sim 6$ Myr} and \mbox{$\sim 30$ Myr}, corresponding to the most and least massive explosion progenitors, respectively. This differs from the previous prescription in the high-mass end in that the reduced explodability delays the onset of CCSNe by $\sim 1$--2 Myr.
Fig. \ref{fig:lifetimes} visualizes the realized lifetimes from \boost{} at \mbox{$Z=0.1$ \zdot}, CCSN explosion energies and stellar metal-enriched mass return from CCSNe and winds in a simulation that employs the full \textit{M25, variable E} model. The first CCSNe that in the \texttt{Original} model occur at $\sim 4.5$ Myr with \mbox{1 B} energy shift to 1--2 Myr later when we account for the metal-dependent explodability of the CO cores. The first CCSNe, while delayed, tend to be more energetic in the \texttt{M25, variable E} model. This is especially the case when the progenitor star has low metallicity.

For a direct comparison of the \texttt{M25} and the \texttt{Original} implementation of CCSN explosions, in Fig. \ref{fig:explosions} right panel we also show the realized CCSN energies as a function of $M_\mathrm{ZAMS}$ in the dwarf galaxy simulation.

\subsection{Electron-capture supernovae}
\label{section:ecsn}
The least massive stars to explode are those that exceed the core mass of about  1.37 M$_{\odot}$ \citep{Nomoto1987,Posdialowski2004,Takahashi2013} for gravitational collapse of the degeneracy-supported oxygen-neon (ONe) core. Super-AGB stars with ONe cores exceeding this limit but remaining cool enough not to ignite core-neon burning end their lives with explosive NS formation in SNe initiated by the electron-capture (EC) onto magnesium and neon in the core \citep{Nomoto1980}. Stars with more massive cores smoothly transition from the ECSN to the iron-core collapse SN fates. 

Over the lower-mass end of CCSN progenitors, $M_\mathrm{CO}$ typically is only marginally more massive than the ONe core mass. Therefore, the former can also be used as a proxy for mapping out the $M_\mathrm{ZAMS}$-threshold, $M_\mathrm{ZAMS,SN}^\mathrm{min}(Z)$ demarcating the transition between non-explosive white dwarf formation and explosive NS formation via ECSNe, at each $Z$ considered. 
The lightest $M_\mathrm{CO}$ in the \boost{} models reach only down to $\sim1.5$ \mdot{} for $M_\mathrm{ZAMS}\sim9$ \mdot{} at our adopted initial $Z=Z_\odot/10$. Therefore, we cannot rigorously determine $M_\mathrm{ZAMS,SN}^\mathrm{min}(Z)$ and its $Z$-dependence. For simplicity, here we adopt a fixed value of $M_\mathrm{ZAMS,SN}^\mathrm{min}(Z)=8$ \mdot{} for all our CCSN model variations but point out that $M_\mathrm{ZAMS,SN}^\mathrm{min}(Z)$ likely increases with increasing $Z$. At lower $Z$, winds are weaker and thus cores grow more massive, compared to the same $M_\mathrm{ZAMS}$ of a higher-$Z$ star. This has important consequences for the resulting SN statistic due to the IMF. We defer a more elaborated determination of $M_\mathrm{ZAMS,SN}^\mathrm{min}(Z)$ to a separate study, only highlighting its relevance here.

\subsection{Initial conditions and simulation parameters}

To test the impact of $M_\mathrm{CO}$ and metal-dependent explodability on the long-term baryon cycle of dwarf galaxies, we run a set of idealized dwarf galaxy models with various physical inputs. The initial condition is always the same Wolf–Lundmark–Melotte like galaxy in compact configuration as described in \citet{2017MNRAS.471.2151H}. The galaxy has a virial mass of \mbox{$4\times 10^{10}$ \mdot} and includes a \mbox{$4\times 10^7$ \mdot{}} gas disk and a \mbox{$2\times 10^7$ \mdot{}} initial stellar disk. The disk scale length is \mbox{0.73 kpc}. The gas and stellar mass resolution is \mbox{$\gtrsim4$ \mdot{}} and the dark matter has a particle mass resolution of \mbox{$\sim6800$ \mdot{}}. The force-softening length is \mbox{62 pc} for dark matter and \mbox{0.1 pc} for baryonic particles.

We ran in total five dwarf galaxy simulations for the duration of \mbox{1 Gyr} while varying some of the feedback input parameters as described in Table \ref{table:parameters}. In addition to the four $E_\mathrm{expl}$ variations, we ran a fifth simulation with the \texttt{Original} setup but without photoionizing stellar radiation, labeled \texttt{Original, no PI}. As shown in \citet{2017MNRAS.471.2151H} and \citet{2021MNRAS.506.3882S}, disabling photoionization leads to more bursty outflows from the dwarf galaxy. Changes due to the other model variations are more subtle or even indistinguishable, depending on the inspected quantity. Table \ref{table:parameters} lists the parameter variations in the five models.

\begin{figure*}
\includegraphics[width=0.9\textwidth]{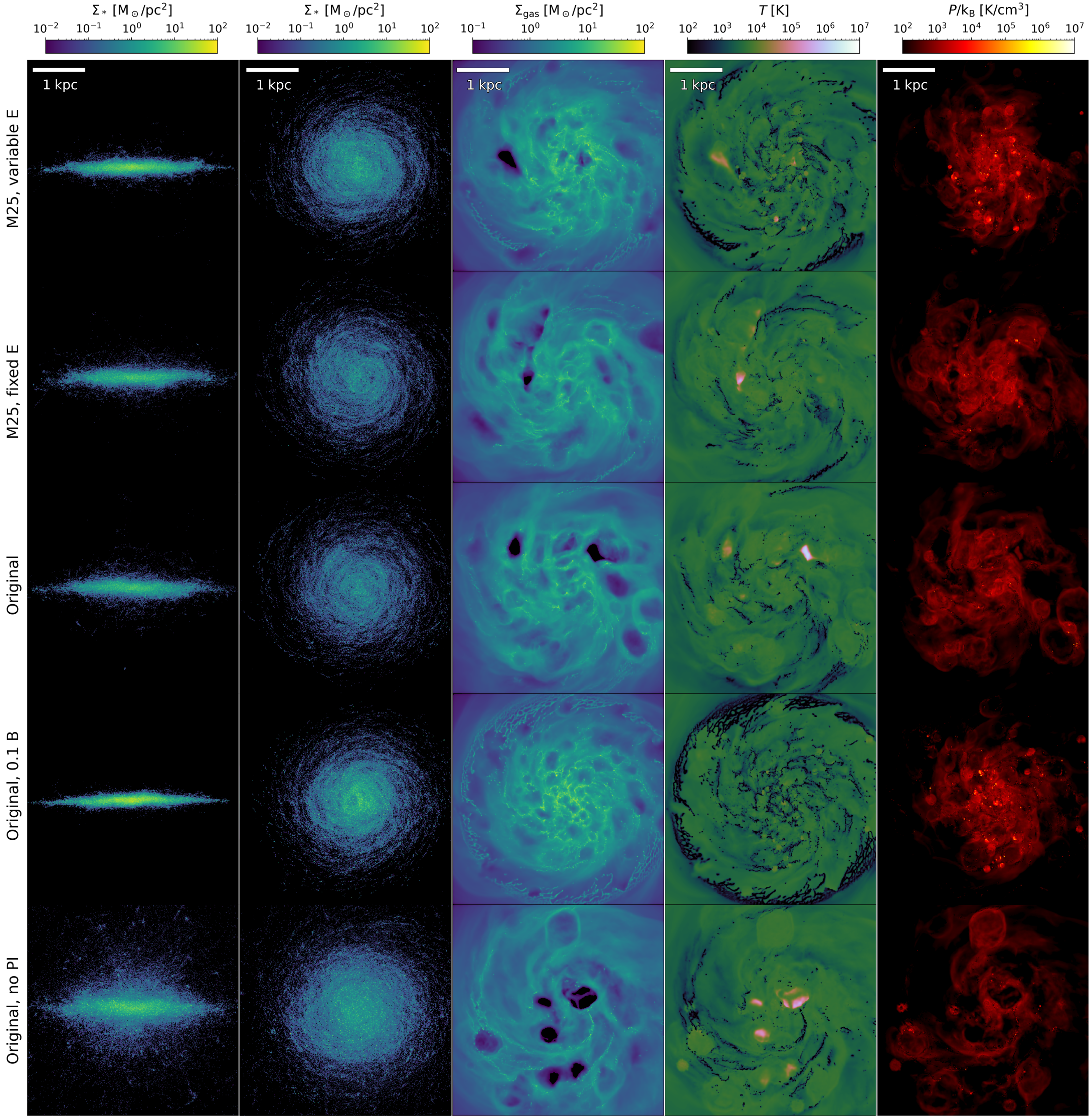}
\caption{Projected maps of stellar surface density in edge-on and face-on projections, gas surface density, gas temperature and gas thermal pressure in face-on projection (panels from \textit{left} to \textit{right}). The simulations are shown at time $\sim 1$ Gyr, one simulation per row: \texttt{M25, variable E}; \texttt{M25, fixed E}; \texttt{Original}; \texttt{Original, 0.1 B}; and \texttt{Original, no PI} from \textit{top} to \textit{bottom}. The time-averaged phase structure is quantified further in Fig. \ref{fig:phasediagram}. \label{fig:collage}}
\end{figure*}

\subsection{Cluster selection}

We inspect the populations of young star clusters that form in the five simulations to quantify how the changes in star formation and outflows translate into changes in star cluster properties. Due to the limited mass resolution, universal gravitational softening and the exclusion of a higher order method to solve the collisional gravitational dynamics, the long-term evolution of star clusters is not captured well in the present models as shown in \citet{2025MNRAS.538.2129L}. Based on \citet{2025MNRAS.538.2129L}, the youngest star clusters should, however, show relatively similar mass functions and star cluster formation efficiencies regardless of the dynamical integration technique. We therefore concentrate here particularly on the cluster demographics right after their formation. Similar to \citet{2020ApJ...891....2L} and \citet{2023MNRAS.522.3092L}, we characterize the young star cluster population using gravitationally bound stellar objects as identified with the \textsc{friends-of-friends} (\textsc{fof}) and \textsc{subfind} \citep{2001MNRAS.328..726S, 2009MNRAS.399..497D} algorithms in \gadget. We use a \textsc{fof} linking length of \mbox{1 pc} and require at least 50 members ($\sim 200$ \mdot) to be gravitationally bound after the unbinding step of \textsc{subfind}. We consider as young clusters those that have mass-averaged stellar age of less than \mbox{10 Myr}.

\section{Results}\label{section:results}

\begin{figure*}
\includegraphics[width=\textwidth]{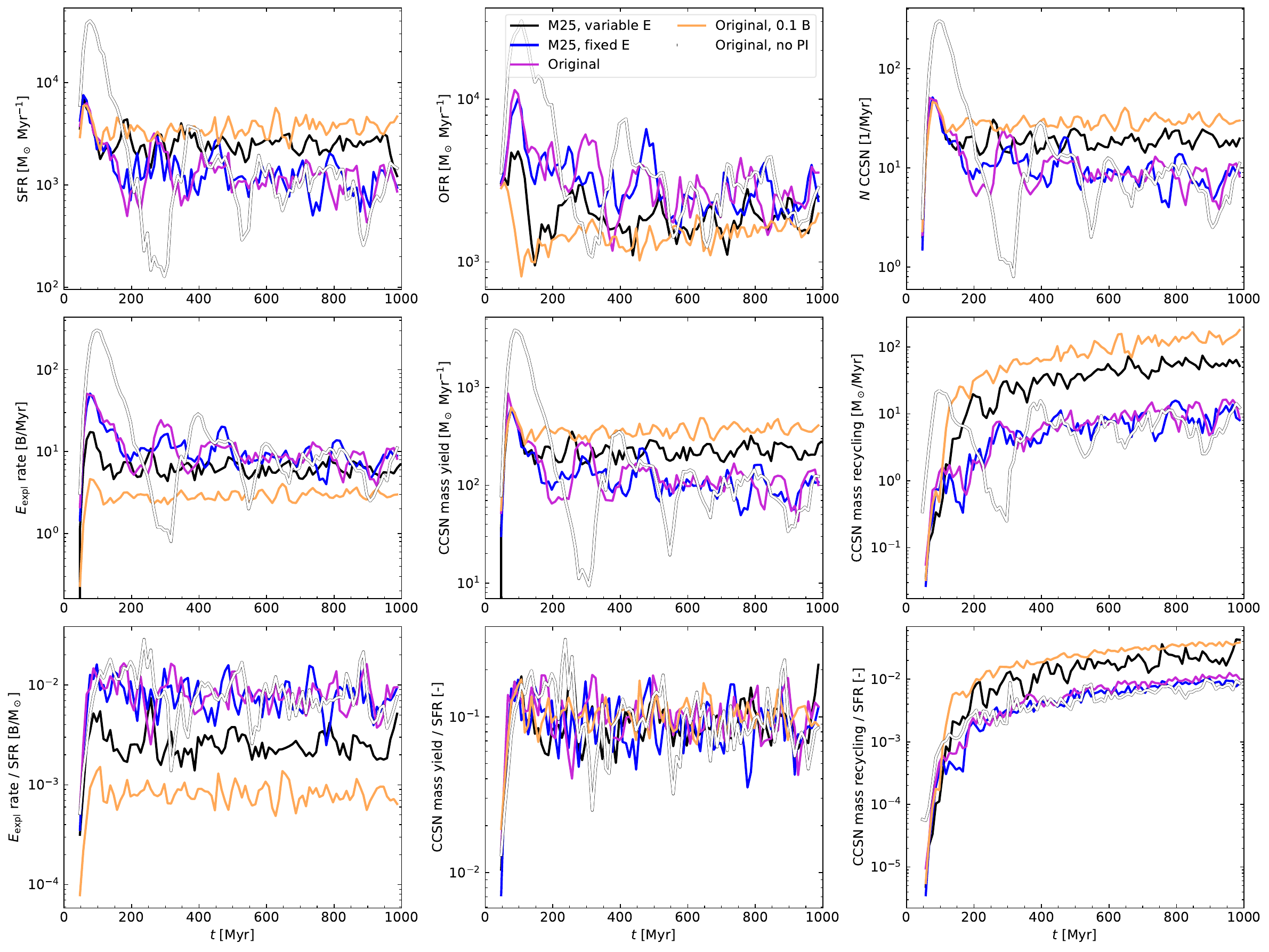}
\caption{Globally integrated properties of the dwarf galaxies run in isolation for \mbox{1 Gyr} (see text for explanations). The line colours from darkest to lightest indicate the simulations run with the full \textit{M25, variable E} explosion model, the \textit{M25} explosion model with a fixed \mbox{1 B} CCSN energy, the \texttt{Original} simulation using the \citet{2023MNRAS.522.3092L} input physics, the \texttt{Original} model with a reduced \mbox{0.1 B} CCSN energy, and the \texttt{Original} model with photoionizing radiation switched off. The dwarf ran with the full \textit{M25, variable E} evolves more as the \mbox{0.1 B} CCSN energy version of \texttt{Original}, whereas the \textit{M25} explodability model with fixed CCSN energy is almost identical to the \texttt{Original} run. All simulations produce approximately the same amount of CCSN enriched material per SFR, however the simulations with reduced CCSN energy recycle a larger fraction of that material. \label{fig:global}}
\end{figure*}

\subsection{BH formation and CCSN explosion statistics}\label{section:explosions}
In the \texttt{M25, variable E} simulation in total 1130 BHs form either through direct-collapse or partial SN ejecta mass fallback over the $M_\mathrm{ZAMS}$-range in-between 8--\mbox{40 \mdot}, in addition to $\sim760$ BHs from stars initially more massive than 40 \mdot. The overall fraction of massive stars ending their lives with BH formation is thus $9.2$\% in our simulation. For comparison, the \texttt{Original} CCSN implementation with fixed limiting $M_\mathrm{ZAMS}=40$ \mdot{} produces a BH fraction of $3.4$\%. The \textit{M25} explodability model thus produces a factor of 2.7 times more BHs. See Section \ref{section:remnants} for more analysis of the remnant population.

In successful CCSNe, including those with fallback-BH outcomes, the realized IMF-averaged energy per SN is now \mbox{0.33 B} in the dwarf galaxy. This is a third of the energy adopted in the \texttt{Original} simulation and in most other modern galaxy simulation studies. Due to the steep IMF at high $M_\mathrm{ZAMS}$, the reduction of the number of SNe per simulation due to reduced explodability is only minor: there are $\sim20200$ CCSNe (including fallback-BH SNe) in the \texttt{M25, variable E} simulation, compared to $21000$ from the run with the \texttt{Original, 1 B} model. The effect of explodability and especially the reduced energies is, however, more enhanced when coupled to the full star formation model in \sphgal{} as discussed in Section \ref{section:global}. Due to variations in the star formation rate between the \texttt{Original} and \texttt{M25, variable E} feedback implementations, there are a factor of 1.8 more CCSNe by number (i.e., a 80\% increase) in the simulation run with the full \textit{M25, variable E} model compared to the \texttt{Original} run.

\subsection{SN feedback on star formation, chemical enrichment and galactic mass outflow dynamics}\label{section:global}

An overview of the simulated dwarf galaxies is given in Fig. \ref{fig:collage} at a simulation time of \mbox{$\sim 1$ Gyr}. By this time all models have passed several cycles of star formation across the disk and settled into a self-regulated state. Fig. \ref{fig:collage} illustrates the variations in both ISM and stellar structure that result from changes to the feedback prescriptions. For instance, largest superbubbles are visible in the simulations that adopt \mbox{$1$ B} for every CCSN, while the \texttt{0.1 B} simulation shows more confined bubble structures and overall slightly colder ISM. Both the \texttt{no PI} and \texttt{0.1 B} simulations form more stars in total compared to the other models, however the \texttt{no PI} run especially develops a visually different disk morphology with a thick stellar component originating from more bursty star formation. The \texttt{no PI} run has a half-mass stellar disk height of $z_{1/2}\sim80$ pc, that is $\sim 80$\% larger compared to the \texttt{Original} run. The \texttt{M25} simulations have approximately the same $z_{1/2}$ as the \texttt{Original} run, while the \texttt{0.1 B} simulation has a $\sim 30$\% thinner stellar disk.

\subsubsection{Star formation, CCSN and outflow rates}

In Fig. \ref{fig:global} we show a variety of globally integrated properties during the entire evolution of the dwarf galaxies in the five simulations. The time-evolution is averaged every \mbox{10 Myr} (\mbox{1 Myr} snapshot interval). In the top two rows we show the star formation rate (SFR), mass outflow rate (OFR), number of CCSN, CCSN energy production, mass yield from CCSNe and recycling of the CCSN material inside new stars, each reported per Myr. The outflow rate has been computed at a height of \mbox{$\pm 1$ kpc} in \mbox{100 pc} slabs as in \citet{2019MNRAS.483.3363H} and \citet{2023MNRAS.522.3092L}. The mass recycling rate measures the cumulative amount of material produced by CCSN that is locked in newly formed stars, essentially quantifying the progress of metal-enriched star formation. In the bottom row, we additionally show the CCSN energy production, mass yield and recycling rates per SFR for a more direct comparison between the different simulations that can have a wide variety of SFRs and hence also CCSN rates. To draw a parallel to the mass loading in outflows discussed later, the SFR scaled quantities can alternatively be thought of as efficiency of energy and mass production by CCSNe, and the recycling rate per SFR as the mass recycling efficiency.

The SFRs in the CCSN model variations increase with decreasing average $E_\mathrm{expl}$, while the OFR decreases with decreasing $E_\mathrm{expl}$. The variability of both rates also decreases with decreasing average $E_\mathrm{expl}$. The number of CCSNe follows the SFR and thus the \texttt{0.1 B} simulation has the highest CCSN production rate out of all simulations. This, however, needs to be convolved with the average $E_\mathrm{expl}$ to produce outflows. Regardless of the large number of CCSNe, the \texttt{0.1 B} model results in the lowest and least variable outflow rates. Disabling photoionization, on the other hand, makes the star formation and outflows very bursty as the stellar winds alone are not efficient enough to regulate the collapse of star forming regions before either all gas is consumed or supernovae step in. In the next Section we show that this results in a number of CCSNe that occur both in orders of magnitude denser and orders of magnitude sparser environments.

\begin{figure}
\includegraphics[width=\columnwidth]{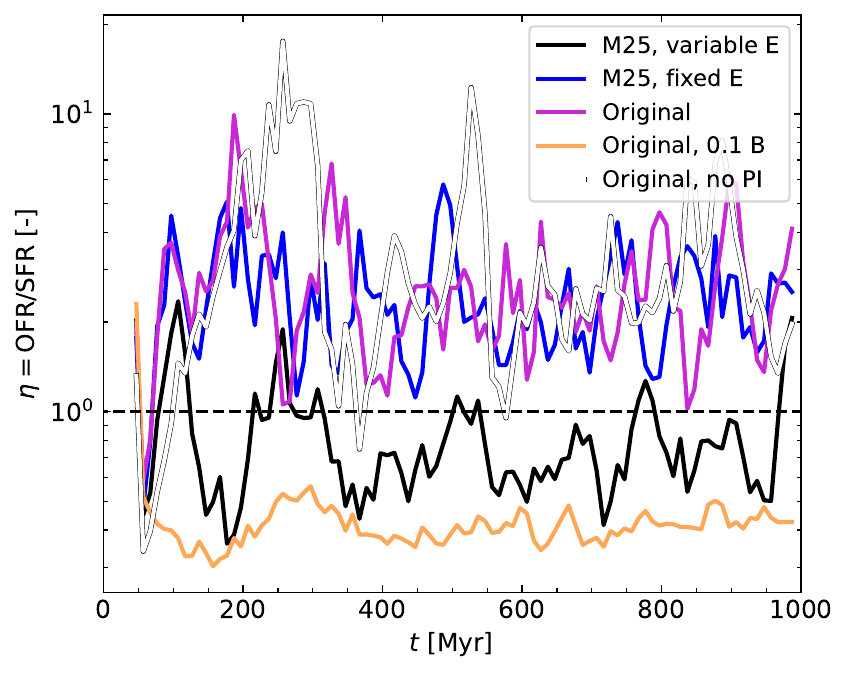}
\caption{The outflow mass loading calculated as OFR per SFR after averaging the rates in 10 Myr timebins. The outflow rate is measured at a height of \mbox{$1$ kpc}. The horizontal dashed line indicates a loading factor of one. The line colours are the same as in Fig. \ref{fig:global}.\label{fig:loadings}}
\end{figure}

The \texttt{M25} simulations and the \texttt{Original} run fall in between these two extremes. When it comes to SN feedback effects on SF, the \texttt{M25, fixed E}-run behaves similar to the \texttt{Original} run. In the \texttt{M25, variable E} run, the average \mbox{$E_\mathrm{expl}=0.33$ B} results in galaxy evolution more akin to the \texttt{0.1 B} simulation but with some burstiness in the outflows. We quantify the mass-loading factors in Fig. \ref{fig:loadings}, defined as SFR per outflow rate. The mass-loading factors (as measured at a height of \mbox{1 kpc}) are of the order of 2 for the \mbox{1 B} simulations and a factor of 2--5 less for the reduced energy simulations. The "canonical" \mbox{$1$ B} simulations therefore remove slightly more gas mass out of the disk in outflows compared to star formation, whereas in the \mbox{$<1$ B} simulations star formation dominates the gas removal. The fountain flow turnover in these kind of simulations is reached approximately at a couple of kpc \citep{2024ApJ...960..100S}, as also indicated by the divergence of loading factors above and below unity at the \mbox{1 kpc} height in Fig. \ref{fig:loadings} in our model variations. Reduction of the CCSN energy thus results in shallower stalling heights for the outflowing material. Inspection of the mass-loading at \mbox{3 kpc} shows that all of the simulations result in sub-unity loading factors at a larger height.

Coming back to Fig. \ref{fig:global}, when scaled with SFR, the total CCSN energies of the \mbox{1 B} simulations match each other on average as expected. The SFR-scaled panels show that while the $E_\mathrm{expl}$ model variations produce on average $\sim 33\%$ and 10\% of the CCSN energy of the \texttt{Original}, the mass injected in CCSNe remains practically unchanged. Since no changes were made to the metal-dependent mass yields per CCSN (apart from the slightly reduced number of CCSN and incidence of fallback-BH formation events in the \textit{M25} models), all the SFR-scaled mass yields agree with each other at an average efficiency of $\sim 10\%$ of the total newly formed stellar mass. 

\begin{figure}
\includegraphics[width=\columnwidth]{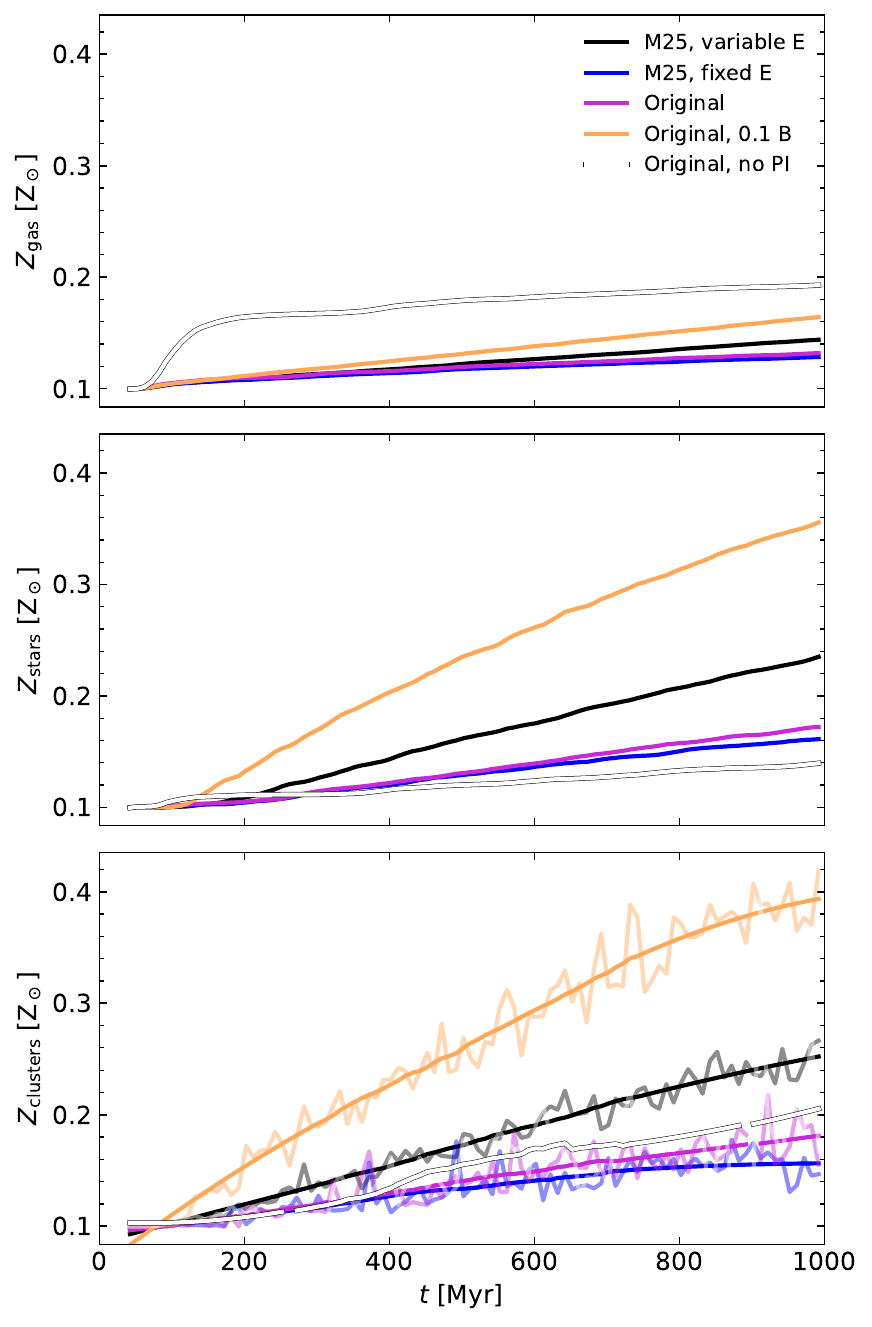}
\caption{The evolution of the mass-averaged metallicity in gas (\textit{top}), stars (\textit{middle}) and stars that are in young (\mbox{$<10$ Myr}) gravitationally bound star clusters (\textit{bottom}). The bottom panel includes both a 10 Myr average and a 50 Myr average. The y-axis range is the same in all panels to visualize the differences in each component and each simulation. The line colours are the same as in Fig. \ref{fig:global}.\label{fig:metals}}
\end{figure}

\subsubsection{Mass-recycling and metal enrichment}

Out of the produced CCSN explosion ejecta mass, a percentage of $\sim 10$--$40\%$ is trapped inside new stars, resulting in gradual increase in stellar metallicities. The fraction increases with decreasing $E_\mathrm{expl}$ in line with the reduced outflows. The recycling rate is by nature a cumulative quantity, therefore its plateau on a Gyr timescale indicates a near equilibrium balance between metal production and removal in outflows in these low-mass galaxies. While the overall ISM evolution of the \texttt{no PI} run is quite different from the other model variations, the resulting CCSN yield-recycling efficiency is still very similar to the other \mbox{1 B} simulations. The periods of enhanced CCSN rate therefore do not lead to enriched star formation, as will be verified next. Rather, the $E_\mathrm{expl}$ controls how much of the produced yield ends up locked in new stars.

In Fig. \ref{fig:metals} we take a closer look at the metallicity evolution.
It allows us to quantify the metal-recycling on a galactic scale by showing the evolution of the average metallicity in the gas, stars and stars in young (\mbox{$<10$ Myr}) gravitationally bound clusters. The gaseous metallicity is enhanced by 25--60\% in the runs with photoionization and by a factor of 2 in the \texttt{no PI} simulation over the timespan of 1 Gyr. While the \texttt{no PI} simulation shows by far the most metal-enriched gas composition as a result of the initial burst of star formation, it is not translated into enriched star formation due to the explosive outflows. In line with this, the recycling rates and stellar metallicities in the simulation with no photoionization shown in Fig. \ref{fig:global} and in Fig. \ref{fig:metals} agree well with the other simulations in which the CCSN energy is fixed at \mbox{1 B}. Once the CCSNe set in, the enhanced outflows compensate for the prolonged star formation episodes and no excess metal-recycling occurs. The \mbox{1 B} simulations result in an average increase of 25--50\% in the stellar metallicity from 0.1 \zdot{} to $\sim0.16$ \zdot. The \texttt{M25, variable E} and \texttt{0.1 B} runs, on the other hand, show factors of 2.2 and 3.5 increase in average stellar metallicity and overall they form the most metal-enriched star clusters. The less energetic CCSNe allow more of the metals to remain in the ISM, to be locked in stars. When the cluster metallicity is contrasted against the globally averaged stellar metallicity, the \texttt{0.1 B} and \texttt{no PI} lock most efficiently metals into star clusters. 

\begin{figure}
\includegraphics[width=0.99\columnwidth]{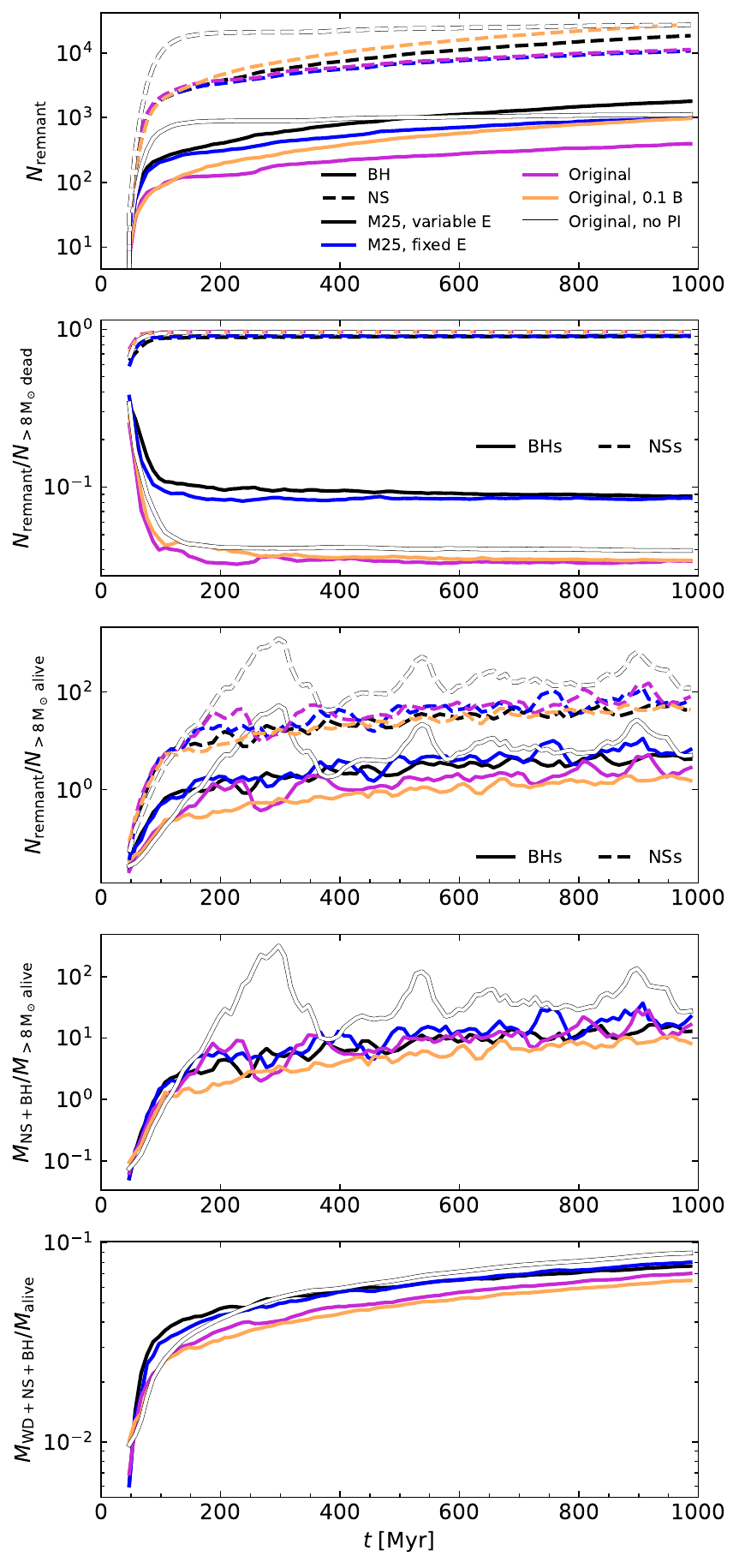}
\caption{The remnant statistics in the five simulations. The panels from \textit{top} to \textit{bottom} are the total number of BHs (solid lines) and NSs (dashed lines), the ratio of BHs and NSs to dead massive ($M_\mathrm{ZAMS}>8$ \mdot) stars, the ratio of BHs and NSs to  massive stars that are alive, the mass ratio of remnants to alive massive stars and the mass ratio of BHs+NSs+WDs divided by the mass in all alive stars. The line colors are the same as in Fig. \ref{fig:global}.\label{fig:fractions}}
\end{figure}

\subsection{Baryon cycle mass loss into compact remnant formation}\label{section:remnants}

Fig. \ref{fig:fractions} shows the overall statistics of BHs and NSs formed along the \mbox{1 Gyr} of galactic evolution. We compute the fraction of BHs and NSs out of all dead massive stars, as well as the number and mass ratios between remnants and alive massive stars to show the ratio of 'dark' to luminous objects. For completeness we also show the mass ratio of all compact objects (formed from progenitor stars with $M_\mathrm{ZAMS}>1$ \mdot) to all stars that are still alive. After evaluating the explodability outcome, we take the final stellar mass (accounting for fallback) as the BH mass for failed CCSNe, while for the NSs and white dwarfs (WDs) we assume fixed fiducial masses of \mbox{1.4 \mdot{}} and \mbox{0.6 \mdot}, respectively.

The number of NSs formed dominates over the number of BHs formed by a factor of $\sim 10$ in the \texttt{M25} simulations. We note that the exact ratio of BHs to NSs depends on the adopted lower $M_\mathrm{ZAMS,SN}^\mathrm{min}(Z)$-limit for successful SN explosion (see Sect. \ref{section:ecsn}). The number of NSs may therefore be overestimated in our model. As noted in \ref{section:explosions}, the overall fraction of BHs among the dead massive stars is 2.7 times higher in the \texttt{M25} models compared to the \texttt{Original} models. The mass fraction of dead massive stars (BHs and NSs) to massive ($> 8 \, M_\odot$) luminous stars crosses unity already at simulation time \mbox{$\sim100$ Myr} in all simulations, and outgrows  it by a factor of 10 after 400--500 Myr of the initial star formation episode. 

The \texttt{no PI} simulation again stands out in the figure when the remnant population is scaled with the luminous one. This is due to the large fluctuations of SFR; when SFR drops sharply, the fractions of dark remnants increase even though the stellar populations are not initially different between the simulations. We verified this by checking that the stellar IMFs are quite similar, except for the upper mass end (\mbox{$m_*>100$ \mdot{}}) being slightly less under-populated in the \texttt{no PI} run due to the increased burstiness. The IMF is never fully sampled near the upper limit of 500 \mdot, but occasionally the enhanced SFR allows for larger reservoirs for random sampling of stars. As a result, the gradual deviation downward from the Kroupa IMF occurs at slightly higher stellar masses in the \texttt{no PI} run. This can be seen in the slightly higher BH ratio in the second panel for \texttt{no PI}. The increased number of BHs is, however, not translated into as large an increase in BH numbers when compared to the new \texttt{M25} implementation, where BHs are allowed to form from stellar progenitors with lower ZAMS masses.

When accounting for lower-mass stars and their remnants (WDs), the long-lived progenitors suppress the ratio of dead stars to alive ones. After billion years of evolution, alive stars that started with $M_\mathrm{ZAMS}>1$ \mdot{} still outweigh compact remnants by a factor of 5--10.

\begin{figure}
\includegraphics[width=\columnwidth]{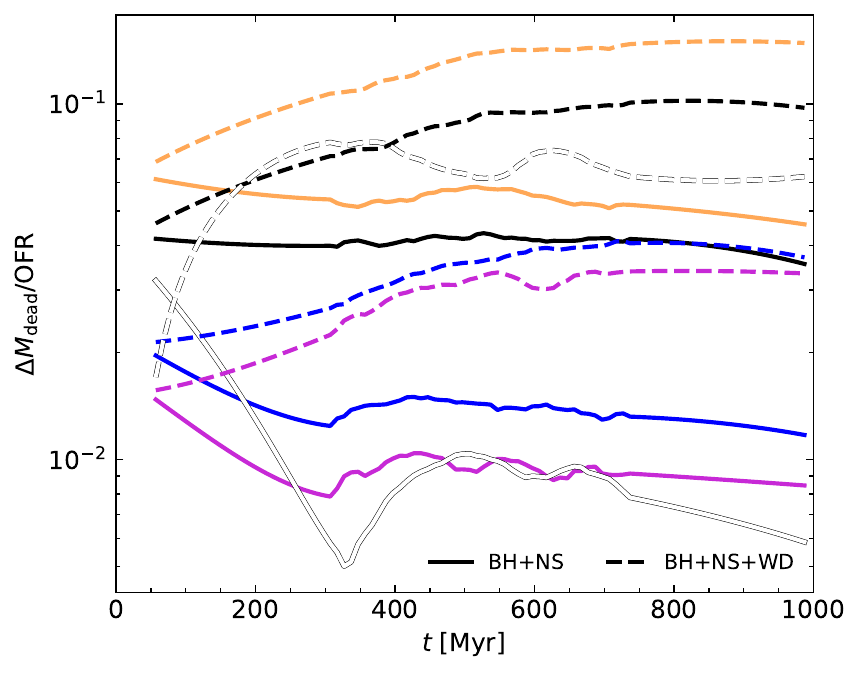}
\caption{The ratio between the mass locked in stellar remnants and gas mass removed in galactic outflows. Remnants from massive progenitors (\mbox{$M_\mathrm{ZAMS}>8$ \mdot}, solid lines) and all compact remnants (\mbox{$M_\mathrm{ZAMS}>1$ \mdot}, dashed lines) are shown separately.  The ratios have been smoothed over a 50 Myr timescale to improve the readability, due to large fluctuations in the resulting values caused mainly by the variability of the outflow rates. The line colours are the same as in Fig. \ref{fig:global}.\label{fig:dead}}
\end{figure}

Fig. \ref{fig:dead} shows the rate of mass locked in massive stellar remnants (BHs and NSs) and overall compact remnants (WDs, BHs and NSs) compared to the galactic outflow rate from Fig. \ref{fig:global}. For every solar mass removed in outflows, 0.1 solar masses is locked up in remnants in the \texttt{M25, variable E} simulation. Of those compact remnants, 40\% by mass are BHs and NSs. The \texttt{0.1 B} simulation is the most efficient in depositing mass into any kind of compact remnants compared to outflows, consistent with the low mass-loading factors seen in Fig. \ref{fig:loadings}. The other simulations are even more outflow-dominated in terms of the rate at which mass is removed from the galactic baryon cycle.

\subsection{Spatial distribution of CCSN explosion sites}

\begin{figure}
\includegraphics[width=\columnwidth]{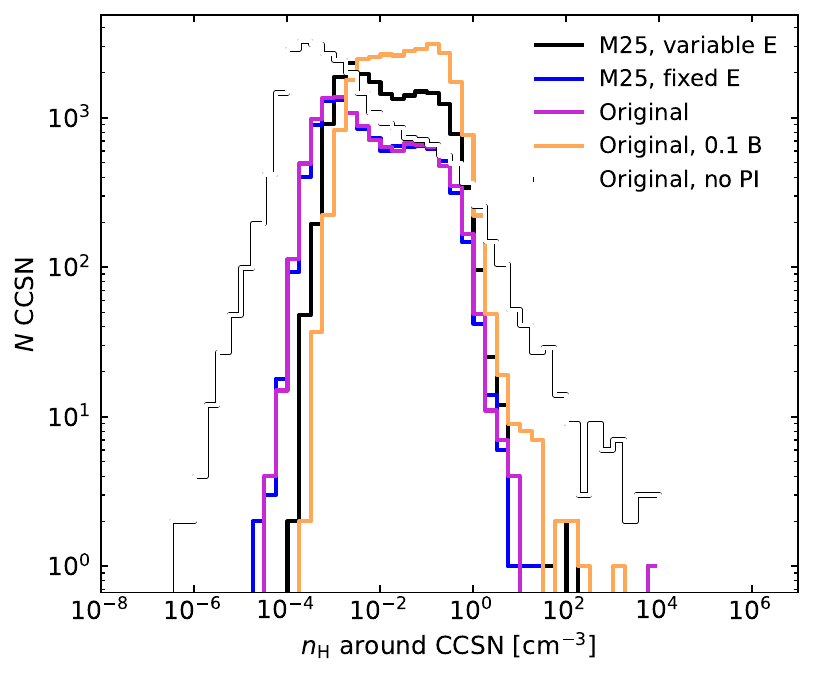}
\caption{The distribution of ambient gas densities at the locations of CCSNe in each of the dwarf galaxy simulations, stacked over \mbox{1 Gyr}. The line colours are the same as in Fig. \ref{fig:global}.\label{fig:SN_density}}
\end{figure}

Fig. \ref{fig:SN_density} shows all the ambient hydrogen number densities, $n_\mathrm{H}$, around CCSNe in each five simulations, stacked across \mbox{1 Gyr} of galactic evolution. The \texttt{Original} simulation shows a two-peaked distribution of ISM densities similar to previous studies \citep{2022MNRAS.509.5938H, 2025MNRAS.538.2129L}. While all of the simulations span at least six orders of magnitude in $n_\mathrm{H}$, the effect of lowered mean-$E_\mathrm{expl}$ in \texttt{M25, variable E} and \texttt{Original, 0.1 B} is to shift $n_\mathrm{H}$ to slightly higher values. The median $n_\mathrm{H}$ in the CCSN model variations spans an order of magnitude from $4\times10^{-3}\,\mathrm{cm}^{-3}$ in the \texttt{Original} and \texttt{M25, fixed E} to $4\times10^{-2}\,\mathrm{cm}^{-3}$ in the \texttt{0.1 B} run. This follows naturally from the phase structure of the ISM (Fig. \ref{fig:phasediagram}), where for the reduced-$E_\mathrm{expl}$ runs the ISM occupies the cool phase for longer. The highest ISM densities around CCSNe in all these model variants match with values observed around evolved massive stars e.g. reported by \citet{2026ApJ..1000...70S}, with the obvious caveat of projection and selection effects as well as different metallicity. 

On the other hand, the \texttt{no PI} run stands out as having a significantly broadened range of explosion environments compared to the other four model variations. Due to the lack of effective early stellar feedback, more CCSNe occur at higher  densities ($n_\mathrm{H}\gtrsim 1 \mathrm{cm}^{-3}$) than in all other runs. In addition, stronger supernova clustering and resulting bursts of feedback in the \texttt{no PI} significantly reduce $n_\mathrm{H}$. The result is an exceptionally large number of CCSNe in low-density environments, and a shift of the median $n_\mathrm{H}$ to $9\times10^{-4}\,\mathrm{cm}^{-3}$ . This result is in agreement with other numerical studies of the star forming ISM using different codes and detailed implementations of star formation and hydrogen photoionisation \citep[e.g.][]{2021MNRAS.504.1039R,2021MNRAS.506.3882S}.  

\begin{figure*}
\includegraphics[width=\textwidth]{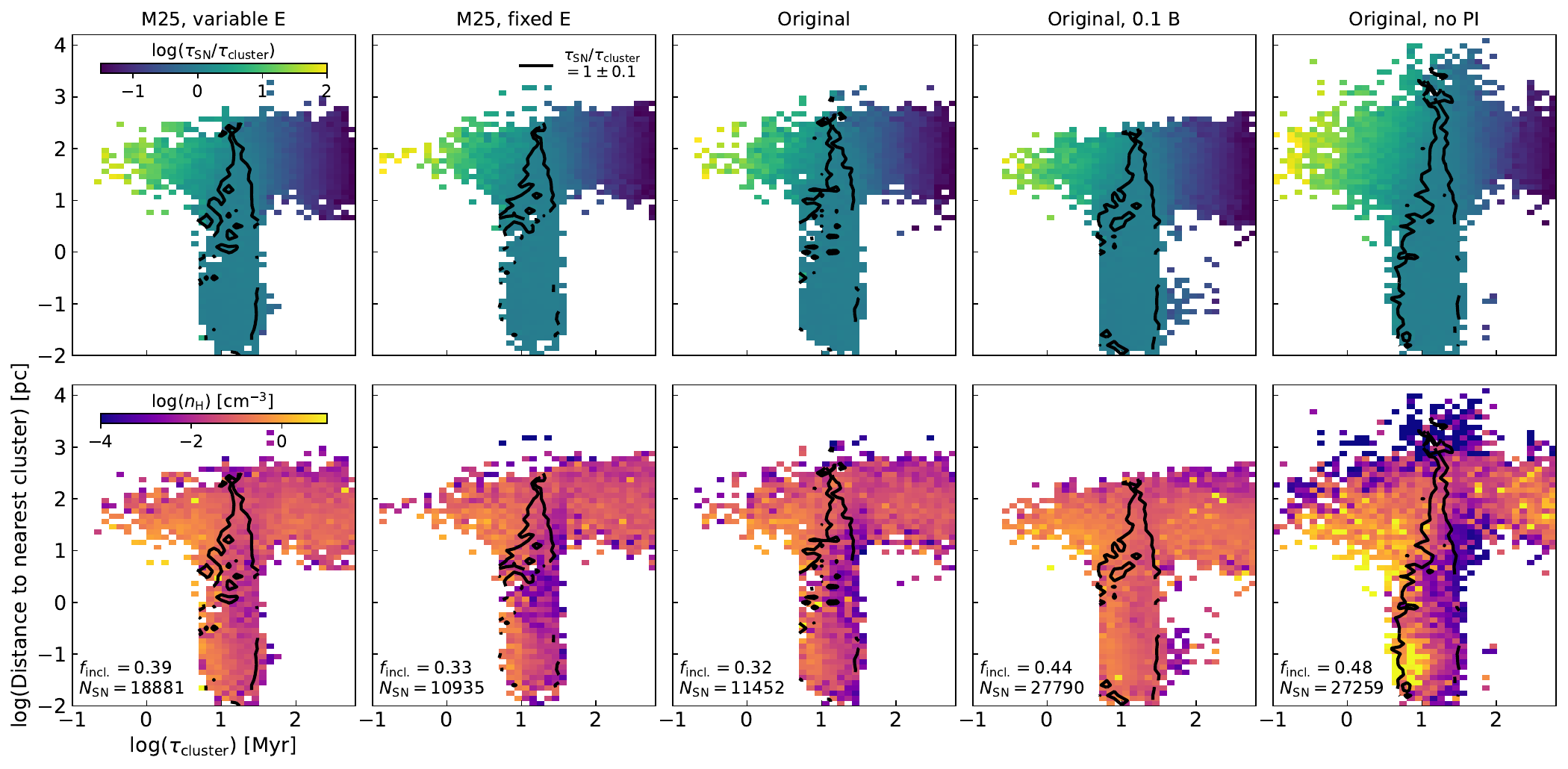}
\caption{The CCSNe in relation to their closest bound star cluster during \mbox{1 Gyr} of evolution. The axes show the binned age and distance of nearest cluster while the colormaps show the average relative age ($\tau\mathrm{SN}/\tau_\mathrm{cluster}$, \textit{top row}; relative age increases from darker to lighter colors) and the average ambient gas density at the location of the CCSNe (\textit{bottom row}; density increases from darker to lighter colors) per bin. The contours indicate the regions where the relative age of the CCSN progenitors and the nearest star cluster is on average $1\pm0.1$. The bottom row also indicates the fractions of CCSNe that are estimated to have occurred within the birth clusters ($f_\mathrm{incl.}$; $\tau_\mathrm{SN}/\tau_\mathrm{cluster}\sim 1\pm0.1$ and distance $<10$ pc) as well as the total number of CCSNe. \label{fig:SN_cluster}}
\end{figure*}

To separate consecutive CCSNe occurring in a given star cluster (i.e. those events potentially directly affected by other physically associated ones) from those that go off in the inter-cluster medium, Fig. \ref{fig:SN_cluster} shows the stacked CCSN events mapped according to the distance and age of the nearest gravitationally bound star cluster. The top row is color coded according the age ratio between the progenitor star and the star cluster, while the bottom row shows $n_\mathrm{H}$ from Fig. \ref{fig:SN_density}. The contours indicate always a relative SN age of $1\pm0.1$, in order to associate CCSN events to their birth star clusters (i.e. those that formed at the same time). The distance axis is an indicator for physical proximity -- together these two quantities provide an impression of CCSNe that are associated with the star clusters in which the CCSN progenitor was born. The fraction of CCSNe that occur close ($<10$ pc) to a similar age cluster ($\tau_\mathrm{SN}/\tau_\mathrm{cluster}\sim 1\pm0.1$) is up to half of all CCSNe in all simulations. The \texttt{no PI} and \texttt{0.1 B} simulations have the most clustered CCSNe (44--48\%) while \texttt{Original} and \texttt{M25, fixed E} have least CCSNe in clusters (32--33\%). The new \textit{M25} implementation falls in between with 39\% of clustered CCSN fraction.

The color gradient in $n_\mathrm{H}$ inside the contours of the bottom row of Fig. \ref{fig:SN_cluster} shows the average progress from first to last CCSN event inside the star clusters. The gradient is stronger in the \mbox{$1$ B} simulations and the \texttt{no PI} run, that is, the later CCSN events in a given cluster occur in lower densities in these runs. We can therefore see that on average the first CCSN event per cluster does more damage in the $1$ B models compared to the reduced CCSN energy runs (\texttt{M25, variable E} and \texttt{Original, 0.1 B}). This is despite the fact that the most powerful CCSN energies in the full \texttt{M25, variable E} model can be up to a factor of two higher. The stochastic nature of the IMF sampling both in terms of timing and stellar masses does not guarantee that the first explosions be from the most massive stars in a star cluster. Due to the added limitation of mass-budget in low mass star clusters, it may happen that by chance only relatively low-mass CCSN progenitors are realized. Based on the \texttt{M25, variable E} data in Fig. \ref{fig:SN_cluster}, this seems to indeed be the case, since otherwise the consecutive CCSNe should occur at lower $n_\mathrm{H}$ than in the \texttt{M25, fixed E} run. The lower outflow rates in the full \texttt{M25, variable E} model in Fig. \ref{fig:global} supports this interpretation.

Nevertheless, the lowest density CCSNe tend to occur in or close to clusters (darker colour pixels are often seen within the region where the SN can be associated with its cluster), where the consecutive feedback events have a better chance to produce rarefied pockets of ISM. In the \texttt{Original, 0.1 B} simulation there are, in addition, a few CCSN events that originate inside older star clusters. These must be cases where the feedback is not strong enough to halt star formation, and the later CCSNe are products of prolonged or episodic star formation in the clusters. The other exception are CCSNe that occur further than \mbox{$\sim 1$ kpc} away from any cluster -- these are stars that explode outside of the galactic disk after they (or their progenitor gas clouds) have been flung out of the galaxy in the aftermath of the first burst of star formation.

The events outside the contours and beyond a few \mbox{$\sim10$ pc} in Fig. \ref{fig:SN_cluster} are inter-cluster SNe. Interestingly enough, the $n_\mathrm{H}$ values of the inter-cluster CCSNe are relatively high (especially so in the \texttt{Original, 0.1 B} run), indicating that the inter-cluster ISM does not have a particularly low density compared to the star clusters when the SNe occur. The exception is the \texttt{no PI} run where $n_\mathrm{H}$ values are clearly the highest when there are relatively young star clusters nearby.

These results are in line with studies that have investigated the impact of massive runaway stars either explicitly trough dynamical ejections \citep{2025MNRAS.538.2129L} or using sub-grid methods for random ejection of massive stars from their birth locations \citep{2020MNRAS.494.3328A, 2023MNRAS.526.1408S}. The previous studies concluded that removing CCSNe from their birth clusters has only a minor effect on the evolution of the host galaxy and practically no impact on the ISM densities where the CCSNe occur \citep{2023MNRAS.526.1408S}. Overall, ours and previous results highlight the non-trivial interplay between the ISM phase structure, clustered star formation and the various flavours and timescales of stellar feedback across the galactic ISM. Such effects would not be captured without a method that treats massive stars as individual objects, as also argued from the point of view of stochasticity of stellar feedback in \citet{2021MNRAS.502.5417S}.

\subsection{SN effects on star clusters}

\begin{figure}
\includegraphics[width=\columnwidth]{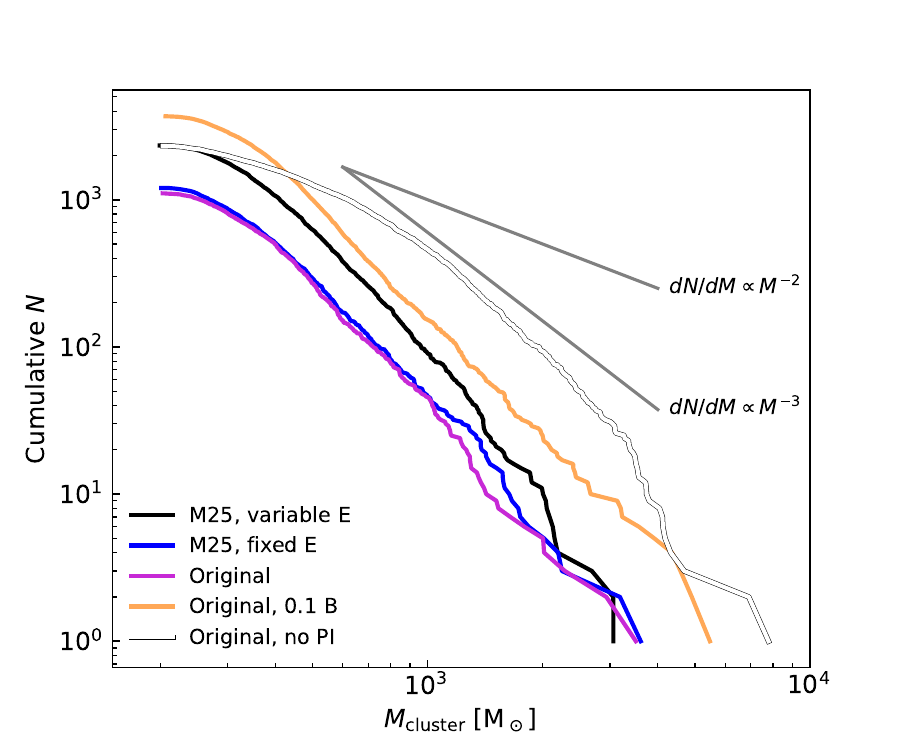}
\caption{The star cluster mass function of young (\mbox{$<10$ Myr}) clusters in the five simulations, stacked over \mbox{1 Gyr} in \mbox{10 Myr} steps. The line colors are the same as in Fig. \ref{fig:global}.\label{fig:CMF}}
\end{figure}

\begin{figure}
\centering
\includegraphics[width=0.7\columnwidth]{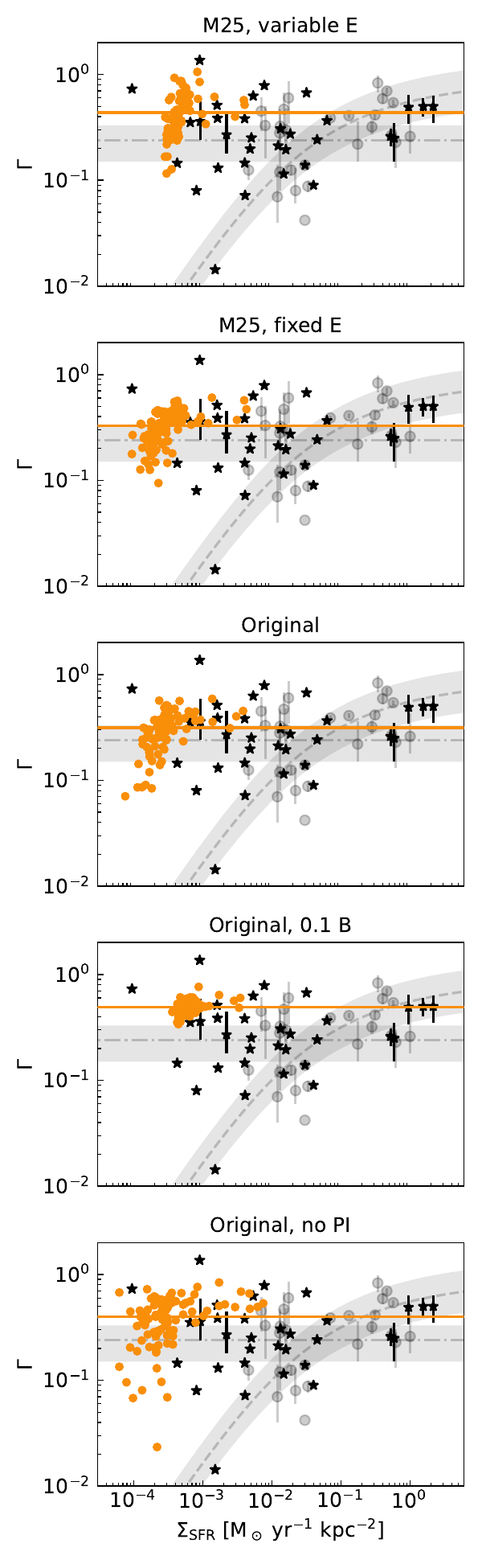}
\caption{The star cluster formation efficiency as a function of SFR surface density in the five simulated dwarf galaxies. The data have been stacked over \mbox{1 Gyr} in steps of 10 Myr, considering only star formation and clusters that formed over the last 10 Myr to match the observed data (see text for details). The observational and theoretical reference data are from \citet{2011MNRAS.417.1904A}, \citet{2011AJ....141..132P}, \citet{2012ApJ...751..100C}, \citet{2017ApJ...849..128C} and \citet{2023MNRAS.519.3749C} for young clusters ($<10$ Myr; highlighted in black); \citet{2010MNRAS.405..857G}, \citet{2011AJ....142..129A},  \citet{2015ApJ...804..123L}, \citet{2015MNRAS.452..246A}, \citet{2016MNRAS.460.2087H}, \citet{2017ApJ...849..128C}, \citet{2018MNRAS.477.1683M}, \citet{2019A&A...628A..60F} and \citet{2020MNRAS.499.3267A} (other results; shown in gray); \citet{2017ApJ...849..128C} (dot-dashed line), and \citet{2012MNRAS.426.3008K} (dashed line); see text for details.\label{fig:CFE}}
\end{figure}

Stellar and SN feedback also leave their imprints on star cluster properties. We quantify this in three ways: by measuring the overall mass function of young star clusters (Fig. \ref{fig:CMF}), the relationship between the star formation rate and the corresponding most massive star cluster (Fig. \ref{fig:Mmax}) as well as the star cluster formation efficiency (Fig. \ref{fig:CFE}).

\subsubsection{Star cluster mass function and most massive cluster}

The star cluster mass function (CMF) is shown in Fig. \ref{fig:CMF} as a cumulative number with decreasing cluster mass $M_\mathrm{cluster}$. A cumulative representation circumvents issues with mass-binning, especially when dealing with low-number statistics such as the steep high-mass end of the CMF. The \texttt{M25} models and the \texttt{Original} simulations show similar, relatively steep CMFs and only vary in normalization that is set by the total number of formed clusters. The \texttt{M25, variable E} simulation hosts relatively fewer massive clusters. The \texttt{0.1 B} simulation has the highest total number of star clusters, more than three times than that of the \texttt{Original} simulation. The \texttt{no PI} simulation has clearly a differently shaped CMF compared to the other models. While the total number of clusters is the same as in the \texttt{M25, variable E} run, the clusters have larger masses and therefore the cumulative CMF starts to plateau at higher cluster mass. The most massive clusters are twice as massive as those in the \texttt{Original} and \texttt{M25} simulations. The behaviour of the CMF shape with changes to the feedback prescription is similar to what was found by \citet{2024A&A...681A..28A} and \citet{2024A&A...691A.231D}. The exclusion of radiative feedback supports the formation of massive star clusters when less energetic channels such as stellar winds are not able to regulate the collapse of the clouds. 

The exact details of the CCSN energetics do not seem to play an important role in controlling the shape of the CMF (apart from the normalization), the main effect is that lower $E_\mathrm{expl}$-runs result in more massive star clusters. The impact of different CCSN models might be particularly relevant in systems with higher star formation rates. Higher SFR density would allow more massive clusters, better guaranteed sampling of high-mass stars, and prolonged gas retainment where the role of CCSN events might be more pronounced. On the other hand, effective early stellar feedback already has a drastic role in halting star formation in star clusters and CCSNe are only needed to finish the process. 

\begin{figure}
\centering
\includegraphics[width=0.8\columnwidth]{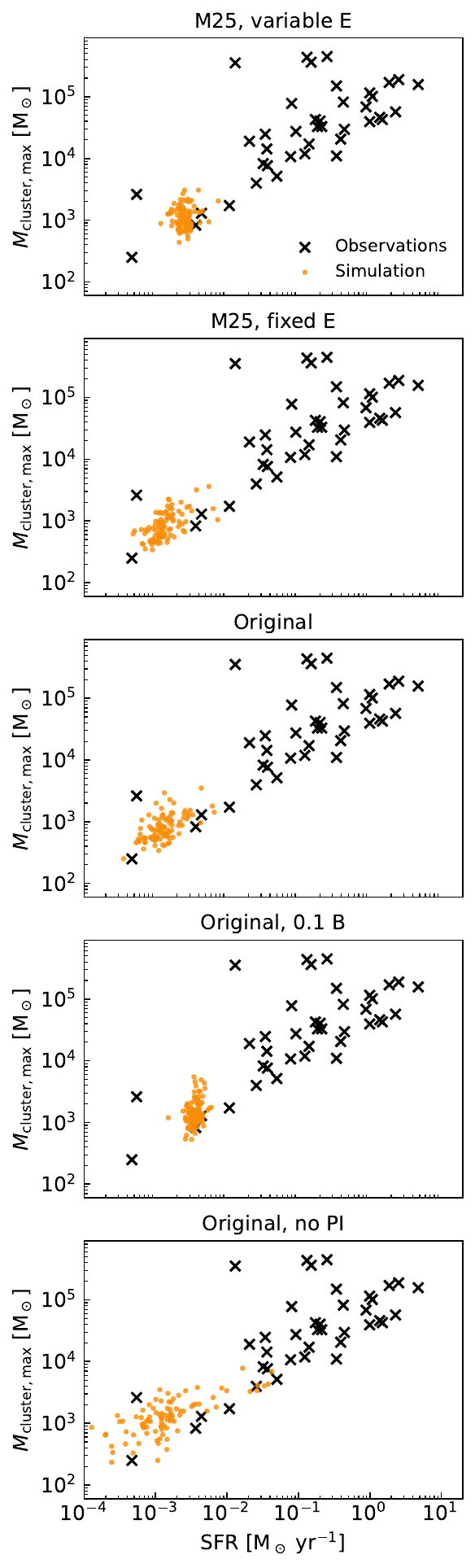}
\caption{The maximum cluster mass as a function of SFR in the five simulated dwarf galaxies. The data have been stacked over \mbox{1 Gyr} in steps of 10 Myr, considering only star formation and clusters that formed within the recent 10 Myr. Observations from \citet{2004MNRAS.348..187W} and \citet{2015A&A...582A..93S} are shown with crosses. \label{fig:Mmax}}
\end{figure}

Fig. \ref{fig:Mmax} shows the mass of the most massive star cluster as a function of the star formation rate, stacked in \mbox{10 Myr} steps across \mbox{1 Gyr}. At face value it seems that the initial star cluster properties are robustly in agreement with observations regardless of the details of the input model. Even without accurate dynamical modelling of the stellar component, the initial properties of the star clusters can be reasonably well captured in such high-resolution simulations. A closer inspection of the individual galaxies however shows that the datapoints for the lowest observed $M_\mathrm{cluster,max}$ also correspond to the only galaxies in this sample with similar gas mass to our simulated galaxy (IC 1613 and Sextans A), i.e. less than $10^8$ \mdot{} \citep{2013AJ....145..101K, 2023AJ....165..260D}. The \texttt{no PI} run seems to span a significantly wider range in SFR than those low-mass galaxies, and incidentally also forms more massive clusters than what is expected from the observed analogues. The other simulations are closer to the expected range of SFRs, and their $M_\mathrm{cluster,max}$ values are either similar or larger than the observed counterparts.

\subsubsection{Star cluster formation efficiency}

Finally, we address the star cluster formation efficiency (CFE, or $\Gamma$), defined as the stellar mass in clusters younger than \mbox{10 Myr} divided by the star formation rate in the same time interval. The \mbox{10 Myr} timespan is matched to the youngest age bins of most observational catalogues. Fig. \ref{fig:CFE} shows the values of CFE measured in \mbox{10 Myr} steps. The observational data in Fig. \ref{fig:CFE} showcases the star cluster formation efficiencies for young clusters ($<10$ Myr) in dwarf galaxies and blue compact galaxies (BCGs), and the data are from \citet{2011MNRAS.417.1904A}, \citet{2011AJ....141..132P}, \citet{2012ApJ...751..100C}, \citet{2017ApJ...849..128C} and \citet[][using $\Sigma_\mathrm{SFR}$ of young resolved stellar populations]{2023MNRAS.519.3749C}. Other observations in the figure are from \citet{2010MNRAS.405..857G}, \citet{2011AJ....142..129A},  \citet{2015ApJ...804..123L}, \citet{2015MNRAS.452..246A}, \citet{2016MNRAS.460.2087H}, \citet{2017ApJ...849..128C}, \citet{2018MNRAS.477.1683M}, \citet[][age limit of $30$ Myr]{2019A&A...628A..60F} and \citet{2020MNRAS.499.3267A}. We also indicate the best fit constant $24\pm9$\% from \citet{2017ApJ...849..128C}, and the analytic model from \citet{2012MNRAS.426.3008K} across a range of $\pm0.2$ dex in CFE.

For all the simulations, the average CFE remains between 31\% and 50\%. There is some stochasticity in the CFE values of all simulations, with \texttt{Original, 0.1 B} being the least scattered dataset. Notably, none of the simulations that include photoionizing radiation show CFE values below $\sim10\%$. Though observational estimates have large uncertainties, these CFE values are within a similar range as in \citet{2023MNRAS.519.3749C} for systems with resolved star formation histories measured for similar stellar and gas mass as ours \citep{2013AJ....145..101K, 2019MNRAS.484.4897C, 2023AJ....165..260D}. The hydrodynamical simulations therefore seem to agree with observational results for young clusters, which indicate that star formation is clustered regardless of star formation environment and feedback implementation. This is in contrast with the analytic model by \citet{2012MNRAS.426.3008K} where the lowest SFR galaxies were broadcasted to form a minority of their stars in clusters. This may be due to dwarfs not being well described as Toomre-stable thin rotating disk galaxies as assumed in the analytic model. The cluster statistics in low-mass systems can also be dominated by single clusters when SFR is very low, deviating the measurements from expected behaviour assuming time-averaged or spatially averaged quantities. Due to this stochasticity, as with the star cluster mass, the formation efficiency in low-SFR systems is not a good quantity for discriminating between the various CCSN input models either. More targeted comparison on small scales, for instance contrasting observed and simulated star-forming regions individually rather than as galactic averages, might be more useful as a follow-up.

\begin{table*}
  \centering
  \setlength{\tabcolsep}{5pt}
  \renewcommand{\arraystretch}{1.1}
  \caption{Time-averaged quantities in the five simulations. Unless stated otherwise, quantities are averaged over $t = 400$--$1000$\,Myr, after the initial transient phase. For ratios of two quantities, the ratio is computed first and then averaged. Values larger (smaller) than in the \texttt{Original} run are highlighted in blue (red). The quantities from top to bottom are (1) SFR; (2) OFR; (3) mass-loading factor;
  (4) $E_\mathrm{expl}$ per unit stellar mass formed;
  (5) median $n_\mathrm{H}$ over the full simulation;
  (6) total number of CCSNe over the simulation ($t = 0$--$1$\,Gyr); (7) CCSN rate;
  (8) mass return rate by CCSN ejecta;
  (9) fraction of CCSNe that occur within 10\,pc of a cluster with relative age $\tau_\mathrm{SN}/\tau_\mathrm{cluster} < 1\pm0.1$, over the full simulation;
  (10) mean CFE over the full simulation;
  (11--13) $Z_\mathrm{gas}$, $Z_\mathrm{stars}$, $Z_\mathrm{clusters}$ measured at $t \approx 1$\,Gyr;
  (14--16) respective metal enrichment rates, given by the slope of a linear least-squares fit to $Z(t)$ over $t = 200$--$1000$\,Myr;
  (17) mean mass of all BHs formed over the full simulation;
(18--20) BH, NS and WD formation rates by number;
  (21--23) rate at which mass is locked into BHs, BHs and NSs, and all three remnant types combined;
  (24--25) ratio of the remnant mass formation rate to the outflow rate, for BHs and NSs (BH+NS) and additionally WDs (BH+NS+WD).}
  \label{tab:averages}
  \begin{tabular}{rllccccc}
    \hline
    & Quantity & Unit & \begin{tabular}{@{}c@{}}\texttt{M25,}\\\texttt{variable E}\end{tabular} & \begin{tabular}{@{}c@{}}\texttt{M25,}\\\texttt{fixed E}\end{tabular} & \texttt{Original} & \begin{tabular}{@{}c@{}}\texttt{Original,}\\\texttt{0.1 B}\end{tabular} & \begin{tabular}{@{}c@{}}\texttt{Original,}\\\texttt{no PI}\end{tabular} \\
    \hline
    (1) & $\langle \mathrm{SFR} \rangle$ & M$_\odot$\,yr$^{-1}$ & \hi $2.5\times10^{-3}$ & $1.2\times10^{-3}$ & $1.2\times10^{-3}$ & \hi $3.7\times10^{-3}$ & \lo $1.1\times10^{-3}$ \\
    (2) & $\langle \mathrm{OFR} \rangle$ & M$_\odot$\,yr$^{-1}$ & \lo $1.8\times10^{-3}$ & $2.7\times10^{-3}$ & $2.7\times10^{-3}$ & \lo $1.5\times10^{-3}$ & \hi $2.9\times10^{-3}$ \\
    (3) & $\langle \eta \rangle$ & -- & \lo 0.76 & \lo 2.4 & 2.5 & \lo 0.41 & \hi 3.2 \\
    (4) & $\langle \dot E_\mathrm{expl}/\mathrm{SFR} \rangle$ & B\,M$_\odot^{-1}$ & \lo $2.6\times10^{-3}$ & \lo $7.8\times10^{-3}$ & $8.1\times10^{-3}$ & \lo $8.3\times10^{-4}$ & $8.1\times10^{-3}$ \\
    (5) & $\tilde n_\mathrm{H}$ & cm$^{-3}$ & \hi $1.1\times10^{-2}$ & \hi $3.9\times10^{-3}$ & $3.6\times10^{-3}$ & \hi $3.6\times10^{-2}$ & \lo $8.8\times10^{-4}$ \\
    \hline
    (6) & $N_\mathrm{SN}$ & -- & \hi $1.9\times10^{4}$ & $1.1\times10^{4}$ & $1.1\times10^{4}$ & \hi $2.8\times10^{4}$ & \hi $2.7\times10^{4}$ \\
    (7) & $\langle \dot N_\mathrm{SN} \rangle$ & yr$^{-1}$ & \hi $1.9\times10^{-5}$ & \lo $8.9\times10^{-6}$ & $9.1\times10^{-6}$ & \hi $3.0\times10^{-5}$ & \lo $8.4\times10^{-6}$ \\
    (8) & $\langle \dot M_\mathrm{SN} \rangle$ & M$_\odot$\,yr$^{-1}$ & \hi $2.3\times10^{-4}$ & \lo $1.0\times10^{-4}$ & $1.1\times10^{-4}$ & \hi $3.7\times10^{-4}$ & \lo $1.0\times10^{-4}$ \\
    (9) & $f_\mathrm{incl.}$ & -- & \hi 0.39 & \hi 0.33 & 0.32 & \hi 0.44 & \hi 0.48 \\
    (10) & $\bar\Gamma$ & -- & \hi 0.44 & \hi 0.33 & 0.31 & \hi 0.49 & \hi 0.40 \\
    \hline
    (11) & $Z_\mathrm{gas}$ & Z$_\odot$ & \hi 0.14 & 0.13 & 0.13 & \hi 0.16 & \hi 0.19 \\
    (12) & $Z_\mathrm{stars}$ & Z$_\odot$ & \hi 0.24 & \lo 0.16 & 0.17 & \hi 0.36 & \lo 0.14 \\
    (13) & $Z_\mathrm{clusters}$ & Z$_\odot$ & \hi 0.25 & \lo 0.16 & 0.18 & \hi 0.39 & \hi 0.21 \\
    (14) & $\dot Z_\mathrm{gas}$ & Z$_\odot$\,Gyr$^{-1}$ & \hi 0.045 & \lo 0.026 & 0.028 & \hi 0.067 & \hi 0.038 \\
    (15) & $\dot Z_\mathrm{stars}$ & Z$_\odot$\,Gyr$^{-1}$ & \hi 0.16 & \lo 0.073 & 0.086 & \hi 0.28 & \lo 0.039 \\
    (16) & $\dot Z_\mathrm{clusters}$ & Z$_\odot$\,Gyr$^{-1}$ & \hi 0.17 & \lo 0.064 & 0.090 & \hi 0.31 & \hi 0.12 \\
    \hline
    (17) & $\bar M_\mathrm{BH}$ & M$_\odot$ & \lo 26.9 & \lo 26.4 & 36.3 & \lo 35.9 & \hi 38.3 \\
    (18) & $\langle \dot N_\mathrm{BH} \rangle$ & yr$^{-1}$ & \hi $1.7\times10^{-6}$ & \hi $8.5\times10^{-7}$ & $3.0\times10^{-7}$ & \hi $1.0\times10^{-6}$ & \lo $2.8\times10^{-7}$ \\
    (19) & $\langle \dot N_\mathrm{NS} \rangle$ & yr$^{-1}$ & \hi $1.9\times10^{-5}$ & \lo $8.8\times10^{-6}$ & $9.1\times10^{-6}$ & \hi $3.0\times10^{-5}$ & \lo $8.5\times10^{-6}$ \\
    (20) & $\langle \dot N_\mathrm{WD} \rangle$ & yr$^{-1}$ & \hi $1.6\times10^{-4}$ & \hi $1.0\times10^{-4}$ & $9.8\times10^{-5}$ & \hi $2.3\times10^{-4}$ & \hi $2.3\times10^{-4}$ \\
    (21) & $\langle \dot M_\mathrm{BH} \rangle$ & M$_\odot$\,yr$^{-1}$ & \hi $4.7\times10^{-5}$ & \hi $2.2\times10^{-5}$ & $1.1\times10^{-5}$ & \hi $3.7\times10^{-5}$ & \lo $1.0\times10^{-5}$ \\
    (22) & $\langle \dot M_\mathrm{BH+NS} \rangle$ & M$_\odot$\,yr$^{-1}$ & \hi $7.3\times10^{-5}$ & \hi $3.5\times10^{-5}$ & $2.4\times10^{-5}$ & \hi $7.9\times10^{-5}$ & \lo $2.2\times10^{-5}$ \\
    (23) & $\langle \dot M_\mathrm{BH+NS+WD} \rangle$ & M$_\odot$\,yr$^{-1}$ & \hi $1.7\times10^{-4}$ & \hi $9.5\times10^{-5}$ & $8.3\times10^{-5}$ & \hi $2.2\times10^{-4}$ & \hi $1.6\times10^{-4}$ \\
    (24) & $\langle \Delta M_\mathrm{dead}/\mathrm{OFR} \rangle_\mathrm{BH+NS}$ & -- & \hi 0.041 & \hi 0.013 & 0.0092 & \hi 0.053 & \lo 0.0083 \\
    (25) & $\langle \Delta M_\mathrm{dead}/\mathrm{OFR} \rangle_\mathrm{BH+NS+WD}$ & -- & \hi 0.096 & \hi 0.038 & 0.033 & \hi 0.14 & \hi 0.065 \\
    \hline
  \end{tabular}
\end{table*}

\section{Discussion}
\label{section:discussion}

Table \ref{tab:averages} summarizes the quantities obtained in this study as averages over time either over the entire simulation or after the first initial burst of star formation. For clarity we indicate with colours the values that are above or below the value obtained in the \texttt{Original} simulation. For most of the average quantities (SFR, remnant production, metal enrichment, cluster formation efficiency), the sub-1 B simulations \texttt{M25, variable E} and \texttt{Original, 0.1 B} exceed the \texttt{Original} model. The only quantities that stand out as having lower values are the OFR, the mass-loading and the average $E_\mathrm{expl}$. \texttt{M25, fixed E} is very similar to \texttt{Original}, with slightly enhanced remnant production and reduced chemical enrichment. The \texttt{no PI} run is a mixture: outflows, clustered star formation and enrichment of the gaseous component are enhanced, while SFR, SN rate and enriched star formation are reduced.

\subsection{Chemical enrichment}

While there are uncertainties in the absolute metallicities in observations, dwarf galaxies are observed to host star clusters for which the metallicity increases systematically over time. Recent examples for instance from \textit{EUCLID} report an evolution of the mean cluster metallicity in two low-mass galaxies IC10 and NGC 6822 from \mbox{0.2 \zdot{}} to 0.3--\mbox{0.4 \zdot{}} and from \mbox{0.1 \zdot{}} to \mbox{0.2 \zdot{}} over the past \mbox{1 Gyr} \citep{2026A&A...706A.185H}. The former is a starburst dwarf galaxy, thus the enrichment can be expected to have been more intense in the recent past, while the latter is a more quiescent irregular galaxy with SFR$\sim 0.02$ \mdot{} yr$^{-1}$. The observed cluster-to-cluster metallicities show variations by factors of several across the recent \mbox{1 Gyr} or so, and this is true even in the well-resolved close-by Small Magellanic Clouds \citep{2021A&A...647A.135N, 2025A&A...701A.222S}. The canonical \mbox{1 B} simulations might result in too inefficient chemical enrichment, especially in the case of the \texttt{no PI} simulation, when compared to the range of metallicities observed in low-mass galaxies. Note also that since we do not account for pristine inflows into the galaxy, our enrichment estimates can be considered as an upper limit. Details of chemical enrichment on an element-by-element level would provide an additional constraint for the chemical enrichment. 

The CCSN yields used in this study are not calibrated to the variable $E_\mathrm{expl}$ obtained from \textit{M25}, therefore we omit an abundance analysis in this study. \citet{2026A&A...707A.112A} accounted for the limited explodability by restricting CCSNe only to $M_\mathrm{ZAMS}<25$\,\mdot{} stars, while \cite{2021MNRAS.501.5597G} also included variable $E_\mathrm{expl}$ and mass yields calculated at solar metallicity. The simulations by \cite{2021MNRAS.501.5597G} were shown in \citet{2024ApJ...971..103G} to produce large spreads in overall metal content in bound star clusters. It is however difficult to disentangle the effect of variable $E_\mathrm{expl}$ from the missing pre-supernova feedback prescription in that study, as we have shown here that \texttt{no PI} case produces especially high SFRs and gaseous metallicities compared to the other simulations. We therefore note that \citet{2024MNRAS.530..645L}, \citet{2025ApJ...980...62M} and \citet{2026A&A...707A.112A}, among others, have recently investigated the development of chemical abundance features in greater detail in high-resolution hydrodynamical simulations of low-mass galaxy formation. A similar analysis of the abundance features might be an interesting future direction for simulations that consider the effects of variable $E_\mathrm{expl}$ self-consistently.

The onset of CCSNe per each star-forming region is also delayed in the new model by \mbox{1--2 Myr} compared to the original implementation, due to the reduction in initial stellar mass up to which explosions successfully occur. To first order the delay and reduction of CCSNe alone leads to slightly slower chemical enrichment. The combined impact with the variable $E_\mathrm{expl}$ is, however, to enhance the chemical enrichment through increased SFR. A $\gtrsim5$--$6$ Myr delay in the onset of CCSNe, also shown in Fig. \ref{fig:SN_cluster}, would be in line with the timescale of gas-removal in young star clusters found in observations \citep{2026NatAs..10.1179P}. 

\subsection{Outflows}

Our results for galactic outflows are in qualitative agreement with previous dwarf galaxy models (e.g. \citealt{2025ApJ...979...44S}) and show that the less energetic CCSNe result in less efficient outflows \citep{2021MNRAS.501.5597G}. Though the exact values depend on the measurement height, our mass-loading factors for the 1 B and \texttt{M25} simulations are of similar magnitude to those reported in \citet{2025ApJ...979...44S} for their standard \mbox{1 B} and variable energy model as well as the model where randomly selected 40\% of CCSN events fail to deposit energy. Similar to their results, our \texttt{0.1 B} run results in the least efficient outflows.

Our results support low mass-loading factors for implementations of sub-resolution outflow-launching models (e.g. \citealt{2024MNRAS.527.1216S}). Low mass-loadings of unity or less are in tension with lower resolution cosmological simulations that tend to find significantly higher mass-loading factors for low stellar mass halos \citep{2019MNRAS.490.3234N, 2020MNRAS.494.3971M, 2021MNRAS.508.2979P}. Incidentally, both the previous generations of cosmological simulations and more recent updates to feedback physics, such as those presented in \citet{2023MNRAS.519.3154H}, assume a successful injection of \mbox{1 B} for all massive stars up to $M_\mathrm{ZAMS}\sim 100$ \mdot. Approaches where the input feedback models for cosmological simulations are deliberately calibrated against observed galaxy mass functions and galaxy size relations also favour the canonical values for SN injection as providing the best fits \citep{2026MNRAS.548ag300C}. If we trust the 3D simulations of CCSN explosions and the number of observational results showing the successful CCSNe are limited to relatively low-mass progenitors and have sub-Bethe energies, the low-resolution cosmological simulations may therefore inject (on average) too much CCSN energy and starting too early. We finish by pointing out that observations, while challenging due to projection effects and missing gas phases, indicate mass-loading factors that are lower than those found in the low-resolution simulations that adopt the canonical CCSN parameters (e.g. \citealt{2019ApJ...886...74M, 2025MNRAS.544..132S}).

\section{Conclusions}\label{section:conclusions}

We have implemented and tested a new sub-resolution prescription, \textit{M25, variable E}, for the evolutionary outcomes of massive stars in hydrodynamical galaxy simulations. The model predicts whether a massive star of given final  stellar CO core mass, $M_\mathrm{CO}$ and initial metallicity, $Z$, explodes or not, and if it does, with what explosion energy, $E_\mathrm{expl}$. The model provides \sphgal{} with the distribution of explosion energies -- ranging from $< 0.1$ up to $2$ B --  for successful CCSNe, of failed supernovae outcomes that result in direct-collapse BH formation, and also accounts for BH formation via partial fallback of SN ejecta mass. 

We tested the new explosion model in a low-mass, low-metallicity (\mbox{0.1 \zdot}) galaxy simulation against the previously implemented and widely used "canonical" model, \textit{Original}, assuming a fixed $E_\mathrm{expl}=1$ B for all stars having initial masses of 8--40 \mdot. In addition, we ran modifications of these two model classes: one where $E_\mathrm{expl}$ in the \textit{M25} model was fixed to 1 B, and one where $E_\mathrm{expl}$ in the \textit{Original} model was fixed to 0.1 B. The fifth simulation was run with the original implementation of 1 B for 8--40 \mdot{} stars but with stellar photoionizing radiation switched off. These modifications gave us a set of feedback flavours and we have explored their effects onto the baryon cycle in the dwarf galaxy simulation, summarized next.

The new \textit{M25, variable E} CCSN prescription results in 2.7 times more stellar mass BHs being produced when integrated over \mbox{1 Gyr} of star formation. This is due to the 1) addition of a failed explosion window at the intermediate core-masses of the CCSN progenitors, 2) the downward shift of the upper mass limit for successful explosion and 3) the contribution by fallback-BH formation outcomes. The BHs are on average $\sim25$\% less massive in the new model. The average CCSN $E_\mathrm{expl}$ obtained with the full \textit{M25, variable E} model in a low-metallicity dwarf galaxy simulation is \mbox{0.33 B}, which is 33\% of the standard \mbox{1 B} typically assumed in galaxy evolution models. The reduced CCSN energy per SFR results in a higher SFR with less time-variability and reduces the outflow rates, in a similar way as a flat reduction of the fiducial CCSN energy does. The higher SFR leads to more metal-production, which is also recycled more efficiently in metal-enriched star formation while galactic mass outflow rates (OFRs) are reduced.  The resulting low mass-loading factors -- defined as ratios of SFR to OFR --  are therefore below unity with the new CCSN feedback model.
This finding is in better agreement with observations (which themselves have several caveats) compared to lower-resolution cosmological simulations that tend to require large mass-loading factors when the "canonical" ($1$ B) CCSN explosion model is adopted to produce realistic galaxy populations.

The reduction and delay of the CCSN feedback energy furthermore leads to larger cluster formation efficiency in line with the results of \citet{2022MNRAS.509.5938H}. In line with the increased SFR, there are also twice as many star clusters in the reduced $E_\mathrm{expl}$ simulations compared to the standard 1 B run. The fraction of CCSNe closely associated with their birth clusters is also enhanced in those simulations. When comparing the rates of mass removal either in outflows or in compact remnant production, the new model locks 3--4 times more mass in compact remnants compared to the "canonical" implementation. The galactic outflows, however, still dominate the mass removal across models, even though the mass-loading factors are below unity. Overall, the time-variability of the star formation rate and the scatter in the related efficiencies decrease with decreasing CCSN energy. The simulation without photoionization leads to least enriched star formation, despite the most metal-enhanced ISM. The SFR range and most massive star cluster in that simulation deviate from observed measurements for galaxies at similar stellar and gaseous mass. 

Within each simulation, the ambient CCSN environment densities show similar average values regardless whether they occur in or outside of clusters. This was quantified by measuring the average ambient SN density as a function of distance and age of the nearest star cluster, to associate CCSNe to their birth clusters. The exception are CCSN events that occur in clusters after the previous SN events have locally lowered the ISM densities (or completely outside the galactic disk). In other words, the explosion of CCSNe in lowest-density environments occur in or near somewhat evolved star clusters where feedback has been effective in creating lower density bubbles. The bubbles are more confined in models that have a lower-than-1 B average CCSN energy, with the slightly cooler and denser ISM resulting in a shift to higher ambient densities around CCSNe in those models.

Constrained cloud scale simulations \citep{2025ApJ...990..112C} have already demonstrated that binarity in massive stars increases the radiation output and can therefore have a drastic impact on the star-forming ISM on scales of star clusters. In the absence of direct methods for treating binary stars in the \sphgal{} code, we have excluded the explodability estimates for stars that have undergone mass transfer due to interactions with a binary companion. \textit{M25} provides the explodability criteria for stars that have undergone interactions with binary companion stars their lives, acting as donors or accretors of mass transfer. The removal of the stellar envelope affects which stars can successfully explode, hence binarity also directly impacts the CCSN-driven aspects of the global baryon cycle. Future developments to state-of-the-art galaxy evolution models should therefore account for the complexity introduced by stellar evolution in binaries. Our future aim is to address this by combining the built-in gravitational dynamics integrator \ketju{} with fast models for binary evolution typically employed in $N$-body simulations, such as \textsc{SeBa} \citep{2012A&A...546A..70T} or \textsc{sevn} \citep{2023MNRAS.524..426I}).

\section*{Acknowledgments}
NL would like to thank Yunwei Deng for inspiring discussions on star cluster simulations. NL was supported by a Gliese Fellowship at the Zentrum f\"ur Astronomie, Universit\"at Heidelberg, Germany, and acknowledges the support by the European Research Council via ERC Starting Grant ILMATAR (no. 101220269). KM is grateful for support by the Max-Planck-Institute for Astronomy. TN and HTJ acknowledge support from the Deutsche Forschungsgemeinschaft (DFG, German Research Foundation) under Germany's Excellence Strategy - EXC-2094 - 390783311 from the DFG Cluster of Excellence "ORIGINS". HTJ is grateful for support by DFG through the Collaborative Research Centre ``Neutrinos and Dark Matter in Astro- and Particle Physics (NDM),'' Grant No.\ SFB-1258-283604770. The computations were carried out on the Orion cluster hosted by The Max Planck Computing and Data Facility (MPCDF) in Garching, Germany.

This research made use of \textsc{python} packages \textsc{scipy} \citep{2020SciPy-NMeth}, \textsc{numpy} \citep{2020NumPy-Array}, \textsc{matplotlib} \citep{Hunter:2007}, \textsc{pygad} \citep{2020MNRAS.496..152R}, \textsc{h5py} \citep{collette_python_hdf5_2014}, and \textsc{astropy}\footnote{http://www.astropy.org} \citep{2022ApJ...935..167A}. Claude Code Opus 5 and 5.5 by Anthropic were used to polish and optimize the creation of figures and tables.

\section*{Data availability statement}
The data relevant to this article will be shared on reasonable request to the corresponding author. A Jupyter Notebook script implementing the \textit{M25} CCSN model is publicly available via \url{https://zenodo.org/records/15350140}.


\bibliographystyle{mnras}
\interlinepenalty=10000


\appendix\label{section:appendix1}

\section{Time-averaged phasediagram}

\begin{figure*}
\includegraphics[width=\textwidth]{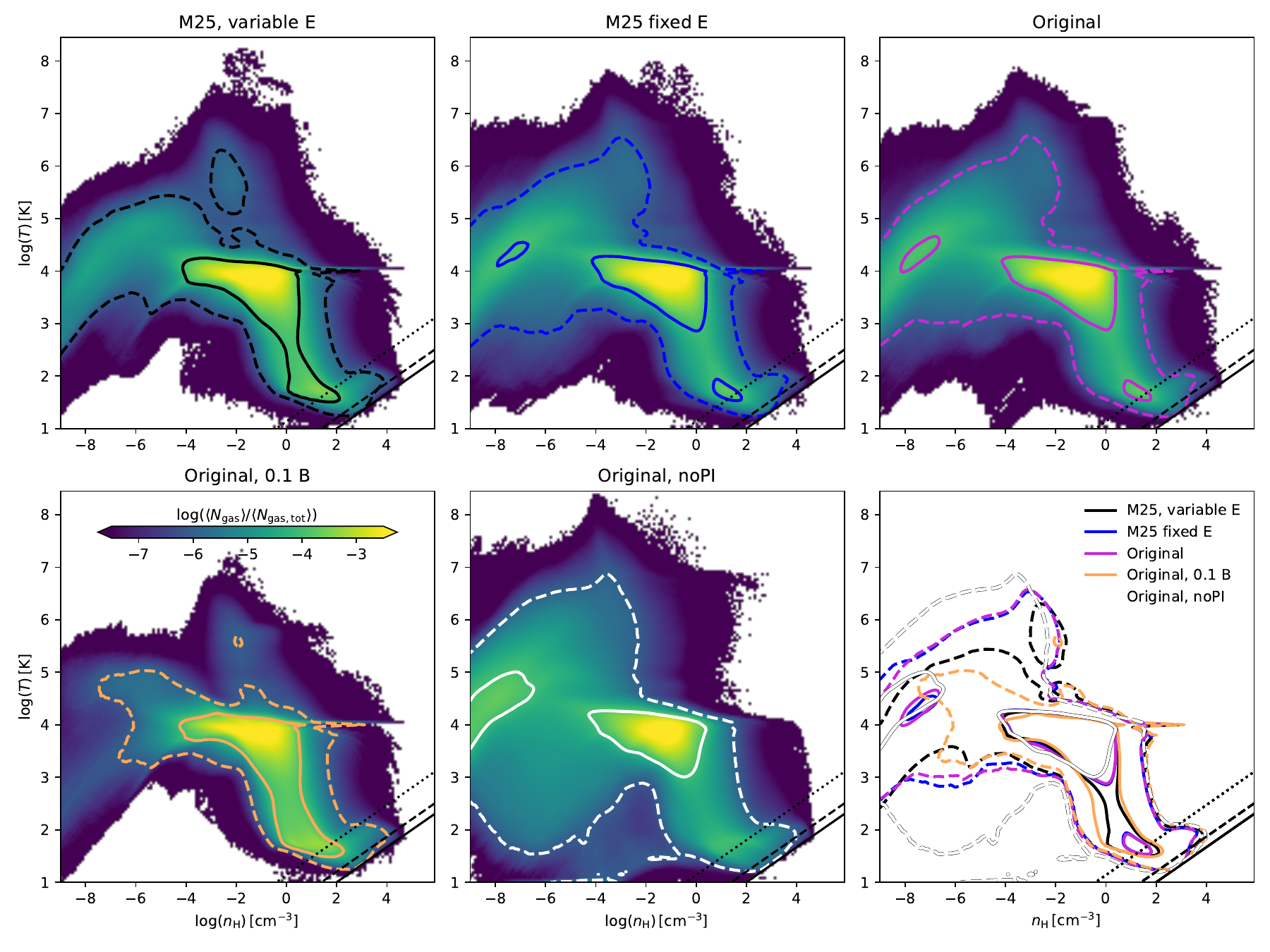}
\caption{The phasediagram of each dwarf galaxy simulation averaged over \mbox{1 Gyr} in \mbox{1 Myr} steps. The colormaps have been normalized to the average number of gas particles per set. The contours indicate the levels of $10^{-6}$ and $10^{-4}$ ($\sim10$ and $\sim1000$ particles per cell) and the line colours are the same as in Fig. \ref{fig:global}. The diagonal lines indicate equivalent values of Jeans mass of 8, 1 and 0.5 SPH kernel masses, wherein the latter is the adopted star formation threshold.\label{fig:phasediagram}}
\end{figure*}

The direct effects of different implementations of stellar feedback can be visualized in a phasediagram. Fig. \ref{fig:phasediagram} shows the time-averaged diagrams of hydrogen number density and gas temperature for each five simulations, computed in 1 Myr steps across \mbox{1 Gyr}. The two contours in the five first panels highlight the different shapes of the binned particle distributions, and the countours have been collected in the last panel for ease of comparison. 

As in Fig. \ref{fig:global}, the \texttt{0.1 B} and \texttt{no PI} runs differ the most from the \texttt{Original} simulations, with the \texttt{M25} model variations sharing features mainly between the \texttt{Original} and the \texttt{0.1 B} models. The reduced hot phase in the lower CCSN energy runs is visible in the diagrams, while the \texttt{no PI} run has large amounts of both hot and cool sparse gas. 


\bsp	
\label{lastpage}
\end{document}